\documentclass[twocolumn,times]{aastex631}
\usepackage{xspace}

\newcommand{\FeKa}{Fe K\ensuremath{\alpha}\xspace}

\newcommand{\NH}{\ensuremath{N_{\mathrm{H}}}\xspace}

\newcommand{\xmm}{{\it{XMM-Newton}}\xspace}
\newcommand{\chandra}{{\it Chandra}\xspace}
\newcommand{\swift}{{\it Swift}\xspace}

\newcommand{\hst}{{HST}\xspace}

\newcommand{\ngc}{{NGC~3516}\xspace}

\newcommand{\nustar}{{\it NuSTAR}\xspace}

\newcommand{\ergflux}{{\ensuremath{\rm{erg\ cm}^{-2}\ \rm{s}^{-1}}}\xspace}
\newcommand{\ergs}{{\ensuremath{\rm{erg\ s}^{-1}}}\xspace}
\newcommand{\cm}{{\ensuremath{\rm{cm}^{-2}}}\xspace}
\newcommand{\den}{{\ensuremath{\rm{cm}^{-3}}}\xspace}

\newcommand{\spex}{\xspace{\tt SPEX}\xspace}
\newcommand{\pion}{\xspace{\tt pion}\xspace}

\accepted{by ApJ on December 12, 2021}

\shorttitle{Changing-look event in NGC 3516: continuum or obscuration variability?}
\shortauthors{Mehdipour et al.}

\graphicspath{{./}{figures/}}
\begin{document}

\title{\large Changing-look event in NGC 3516: continuum or obscuration variability?}

\correspondingauthor{Missagh Mehdipour}
\email{mmehdipour@stsci.edu}

\author[0000-0002-4992-4664]{Missagh Mehdipour}
\affiliation{Space Telescope Science Institute, 3700 San Martin Drive, Baltimore, MD 21218, USA}

\author[0000-0002-2180-8266]{Gerard A. Kriss}
\affiliation{Space Telescope Science Institute, 3700 San Martin Drive, Baltimore, MD 21218, USA}

\author[0000-0003-2663-1954]{Laura W. Brenneman}
\affiliation{Center for Astrophysics | Harvard \& Smithsonian, 60 Garden Street, Cambridge, MA 02138, USA}

\author[0000-0001-8470-749X]{Elisa Costantini}
\affiliation{SRON Netherlands Institute for Space Research, Niels Bohrweg 4, 2333 CA Leiden, the Netherlands}
\affiliation{Anton Pannekoek Institute, University of Amsterdam, Postbus 94249, 1090 GE Amsterdam, The Netherlands}

\author[0000-0001-5540-2822]{Jelle S. Kaastra}
\affiliation{SRON Netherlands Institute for Space Research, Niels Bohrweg 4, 2333 CA Leiden, the Netherlands}
\affiliation{Leiden Observatory, Leiden University, PO Box 9513, 2300 RA Leiden, the Netherlands}

\author[0000-0002-6620-6357]{Graziella Branduardi-Raymont}
\affiliation{Mullard Space Science Laboratory, University College London, Holmbury St. Mary, Dorking, Surrey, RH5 6NT, UK}

\author{Laura Di Gesu}
\affiliation{Italian Space Agency (ASI), Via del Politecnico snc, 00133, Roma, Italy}

\author[0000-0001-5924-8818]{Jacobo Ebrero}
\affiliation{Telespazio UK for the European Space Agency (ESA), European Space Astronomy Centre (ESAC), Camino Bajo del Castillo,s/n, E-28692 Villanueva de la Ca\~nada, Madrid, Spain}

\author[0000-0001-7557-9713]{Junjie Mao}
\affiliation{Department of Physics, Hiroshima University, 1-3-1 Kagamiyama, Higashihiroshima, Hiroshima 739-8526, Japan}
\affiliation{Department of Physics, University of Strathclyde, Glasgow G4 0NG, UK}
\affiliation{SRON Netherlands Institute for Space Research, Niels Bohrweg 4, 2333 CA Leiden, the Netherlands}
%%%%%%%%%%%%%%%%%%%%%%%%%%%%%%%%%%%%%%%%%%%%%%%%%%%%%%%%%%%%%%%%%%%%%%%%%%%%%%%%%%%%%%%%%%%%%%%%%%%%%%%
\begin{abstract}
The Seyfert-1 galaxy NGC 3516 has undergone major spectral changes in recent years. In 2017 we obtained {\it Chandra}, {\it NuSTAR}, and {\it Swift} observations during its new low-flux state. Using these observations we model the spectral energy distribution (SED) and the intrinsic X-ray absorption, and compare the results with those from historical observations taken in 2006. We thereby investigate the effects of the changing-look phenomenon on the accretion-powered radiation and the ionized outflows. Compared to its normal high-flux state in 2006, the intrinsic bolometric luminosity of NGC 3516 was lower by a factor of 4 to 8 during 2017. Our SED modeling shows a significant decline in the luminosity of all the continuum components from the accretion disk and the X-ray source. As a consequence, the reprocessed X-ray emission lines have also become fainter. The {\it Swift} monitoring of NGC 3516 shows remarkable X-ray spectral variability on short (weeks) and long (years) timescales. We investigate whether this variability is driven by obscuration or the intrinsic continuum. We find that the new low-flux spectrum of NGC 3516, and its variability, do not require any new or variable obscuration, and instead can be explained by changes in the ionizing SED that result in lowering of the ionization of the warm-absorber outflows. This in turn induces enhanced X-ray absorption by the warm-absorber outflows, mimicking the presence of new obscuring gas. Using the response of the ionized regions to the SED changes, we place constraints on their density and location.
\end{abstract}
%%%%%%%%%%%%%%%%%%%%%%%%%%%%%%%%%%%%%%%%%%%%%%%%%%%%%%%%%%%%%%%%%%%%%%%%%%%%%%%%%%%%%%%%%%%%%%%%%%%%%%%
%%%%%%%%%%%%%%%%%%%%%%%%%%%%%%%%%%%%%%%%%%%%%%%%%%%%%%%%%%%%%%%%%%%%%%%%%%%%%%%%%%%%%%%%%%%%%%%%%%%%%%%
%%%%%%%%%%%%%%%%%%%%%%%%%%%%%%%%%%%%%%%%%%%%%%%%%%%%%%%%%%%%%%%%%%%%%%%%%%%%%%%%%%%%%%%%%%%%%%%%%%%%%%%
%%%%%%%%%%%%%%%%%%%%%%%%%%%%%%%%%%%%%%%%%%%%%%%%%%%%%%%%%%%%%%%%%%%%%%%%%%%%%%%%%%%%%%%%%%%%%%%%%%%%%%%
\section{Introduction} 
\label{sec:intro}
Outflows from active galactic nuclei (AGN) may couple the supermassive black holes (SMBHs) to their host galaxies. The observed relations between SMBHs and their host galaxies, such as the M-$\sigma$ relation \citep{Ferr00}, indicate that SMBHs and their host galaxies are likely co-evolved through a feedback mechanism. Ionized AGN winds, also referred to as warm-absorber outflows, may play an important role in connecting AGN to their environment (see e.g. \citealt{Laha21}). High-resolution spectroscopy in the UV and X-ray energy bands, in particular with \hst, \chandra, and \xmm, has been instrumental for studying the ionized outflows in AGN. However, many aspects and physical properties of the AGN outflows are still poorly understood. Spectral variability is a hallmark of AGN, yet deciphering its nature and origin remains challenging. As variability can arise from either the accretion-powered radiation or nuclear obscuration/absorption in our line of sight, distinction between these two interpretations is important for advancing our knowledge of the role and impact of outflows in AGN.

\object{NGC 3516} is a notable Seyfert-1 galaxy. Due to its high brightness in both UV and X-rays, and its prominent and clear AGN absorption features, it has been an ideal laboratory for studying the AGN warm-absorber outflows with high-resolution spectroscopy. Over the past few decades there have been many UV and X-ray case studies of the ionized outflows in \ngc, such as publications by \citet{Voit87,Kris96a,Kris96b,Math97,Cos00,Krae02,Net02,Mar08,Tur08,Turn11,Meh10,Holc12,Huer14,Dunn18}. Historically, \ngc has been a remarkably variable AGN, showing changes in both the intrinsic continuum (e.g. \citealt{Meh10,Noda16,Ilic20}) and the intrinsic absorption by the ionized outflows (e.g. \citealt{Cos00,Tur08,Dunn18}). However, over the last few years \ngc underwent a major spectral change, classifying it as a `changing-look AGN' (see e.g. \citealt{Shap19}). 

Changing-look AGN are considered those that alter their appearance from type-1 to type-2 AGN (or vice versa), where the common classification of type-1 and type-2 AGN is based on the optical emission lines, with broad lines predominantly seen in type-1 AGN. Some notable examples of changing-look AGN are \object{Mrk 590} \citep{Denn14}; \object{Mrk 1018} \citep{Cohe86}; \object{NGC 1566} \citep{Okny19}; \object{SDSS J015957.64+003310.5} \citep{LaMa15}; \object{NGC 2617} \citep{Shapp14}; \object{1ES 1927+654} \citep{Trakh19}. Systematic searches using large surveys have revealed an increasing number of other changing-look AGN, see \citet{MacL16,Ruan16,Yang18}. Importantly, major spectral transformations of changing-look AGN are seen in both the intrinsic continuum and the emission lines of the AGN (see e.g. \citealt{LaMa15}). While in general the nature of the changing-look phenomenon may ultimately be attributed to the AGN accretion activity, the physical mechanisms responsible for the transformation are not fully understood. Over the years different scenarios have been postulated for changing-look AGN, such as large changes in the mass accretion rate (e.g. \citealt{Elit14,Runn16,Noda18}); transient nuclear obscuration of the central ionizing source (e.g. \citealt{Tran92,Matt03,Dehg19b}); or in some cases tidal disruption events (e.g. \citealt{Merl15,Komo15}).

\begin{figure*}
\centering
\resizebox{\hsize}{!}{\includegraphics[angle=0]{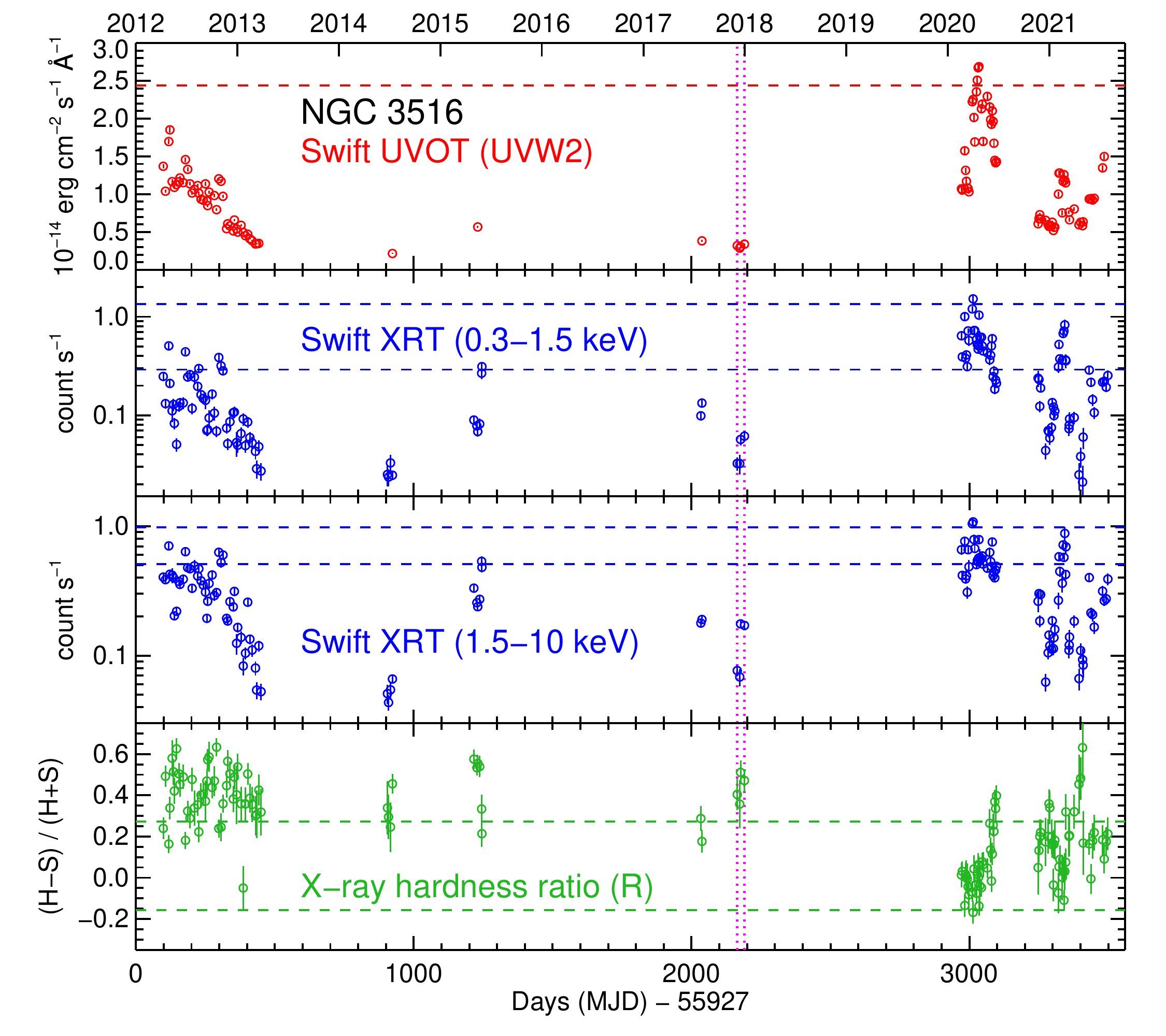}}
\caption{\swift lightcurve of \ngc spanning from January 2012 to July 2021. The bottom panel shows the X-ray hardness ratio ($R$), defined as ${R = (H - S )/(H + S)}$, where $H$ and $S$ are the Swift XRT count rate fluxes in the hard (1.5--10 keV) and soft (0.3--1.5 keV) bands, respectively. The vertical dotted lines in magenta indicate the time of our joint \chandra/LETG and \nustar observations (Obs. 1 and Obs. 2) taken in December 2017 when \ngc was in its low-flux state. The horizontal dashed lines in each panel indicate the range of flux or hardness ratio seen in 2006 during the high-flux state, obtained from the archival \xmm and \swift observations. We use our spectral model fitted to the 2006 \xmm observation (Fig. \ref{fig_fit}) to calculate the equivalent \swift XRT count rates for the purpose of displaying the horizontal lines on this figure. A single \swift/UVOT (UVW2) observation was taken in 2006, and thus only one horizontal line is plotted in the {\it top panel}. The tick marks on the top axis indicate start of each year for reference. The displayed data include the effects of absorption and reddening. Major declines in the UV and X-ray fluxes since 2012 ({\it top and middle panels}), as well as significant X-ray spectral hardening and variability ({\it bottom panels}), are evident in the \swift lightcurve. In 2020 there is a recovery in the flux and the hardness ratio, reaching the 2006 levels, followed by another decline towards the end of the year and further variability in 2021.
\label{fig_lc}}
%\vspace{0.3cm}
\end{figure*}

In the case of \ngc, \citet{Shap19} first reported about its changing look from high-flux to low-flux state. Since then there have been further studies of the changing-look behavior of the optical emission lines from the broad-line region (BLR), and its reverberation mapping, by \citet{Feng21} and \citet{Okny21}. In this paper, by deriving the broadband spectral energy distribution (SED), and photoionization modeling, we aim to determine the intrinsic optical-UV-X-ray continuum and the X-ray absorption in \ngc, before and during the changing-look event. This enables us to examine the two competing scenarios of a change in the accretion-powered radiation versus the appearance of transient obscuring gas. To this end, we obtained two sets of \chandra Low Energy Transmission Grating (LETG) and \nustar observations in December 2017 during the depth of the low-flux state in \ngc. We also obtained {\it Neil Gehrels Swift Observatory} monitoring observations in 2017, taken contemporaneously with \chandra and \nustar, using all the six primary optical and UV filters of the \swift UVOT. This facilitates deriving the optical/UV part of the continuum in our SED modeling. The two sets of \chandra and \nustar observations were separated by about three weeks to probe the observed X-ray spectral variability that was evident in the \swift monitoring of \ngc (see Fig. \ref{fig_lc}). These new 2017 data taken during the low-flux state, and the archival \xmm and \swift data taken in 2006 during the high-flux state, are used for our modeling in this paper to establish the change in the SED and the change in the X-ray absorption in \ngc.

%============================
% FIG: Swift lightcurve
%
%============================

Figure \ref{fig_lc} shows the \swift lightcurve of \ngc from January 2012 to July 2021, revealing the long-term behavior of the changing-look phenomenon. The UV, and soft and hard X-ray fluxes, already appear to decline in 2012 when \swift started to extensively monitor the AGN. For comparison the UV and X-ray flux levels from 2006 are also over-plotted in Fig. \ref{fig_lc} (horizontal dashed lines), when \ngc was in its high-flux state according to archival \swift and \xmm observations (see e.g. \citealt{Tur08,Meh10}). The vertical dotted lines in the \swift lightcurve of Fig. \ref{fig_lc} correspond to when our joint \chandra/LETG and \nustar observations were taken in December 2017 at the low-flux state. The bottom panel of Fig. \ref{fig_lc} shows the X-ray hardness ratio ($R$), defined as ${R = (H - S )/(H + S)}$, where $H$ and $S$ are the \swift XRT count rate fluxes in the hard (1.5--10 keV) and soft (0.3--1.5 keV) bands, respectively. As well as significant spectral hardening in the low-flux state (Fig. \ref{fig_lc}, bottom panel), the hardness ratio $R$ shows remarkable X-ray spectral variability on weeks timescales. The \swift monitoring of \ngc (Fig. \ref{fig_lc}) shows that the UV and X-ray fluxes recovered in 2020, and the hardness ratio reached the soft levels seen during the high-flux state of 2006. \citet{Okny21} have also reported about the re-emergence of \ngc to its high-flux state in 2020. Furthermore, Fig. \ref{fig_lc} shows that towards the end of 2020 there is another decline in flux, as well as significant variability throughout 2021. 

The structure of the paper is as follows. In Section \ref{sect_data} we describe our observations and the data reduction and processing. The modeling of the spectra and the results are presented in Section \ref{sect_model}. We discuss and interpret our findings in Section \ref{sect_discuss}, and give concluding remarks in Section \ref{sect_concl}. Our spectral analysis and modeling were done using the {\tt SPEX} package \citep{Kaa96,Kaas20} v3.06.01. We use C-statistics \citep{Cash79} for fitting the data as it provides the most robust method for X-ray spectral fitting with low count rate statistics; for more information about C-statistics and its implementation in {\tt SPEX} we refer to \citet{Kaas17} and the {\tt SPEX} manual \citep{Kaas20}. The model parameter errors are given at the $1\sigma$ confidence level. We adopt a luminosity distance of 38.1 Mpc (redshift ${z = 0.008836}$, \citealt{Keel96}) in our calculations with the cosmological parameters ${H_{0}=70\ \mathrm{km\ s^{-1}\ Mpc^{-1}}}$, $\Omega_{\Lambda}=0.70$, and $\Omega_{m}=0.30$.

%%%%%%%%%%%%%%%%%%%%%%%%%%%%%%%%%%%%%%%%%%%%%%%%%%%%%%%%%%%%%%%%%%%%%%%%%%%%%%%%%%%%%%%%%%%%%%%%%%%%%%%
%%%%%%%%%%%%%%%%%%%%%%%%%%%%%%%%%%%%%%%%%%%%%%%%%%%%%%%%%%%%%%%%%%%%%%%%%%%%%%%%%%%%%%%%%%%%%%%%%%%%%%%
%%%%%%%%%%%%%%%%%%%%%%%%%%%%%%%%%%%%%%%%%%%%%%%%%%%%%%%%%%%%%%%%%%%%%%%%%%%%%%%%%%%%%%%%%%%%%%%%%%%%%%%
\section{Observations and data reduction} 
\label{sect_data}

\begin{figure}
\hspace{-0.37cm}\resizebox{1.08\hsize}{!}{\includegraphics[angle=0]{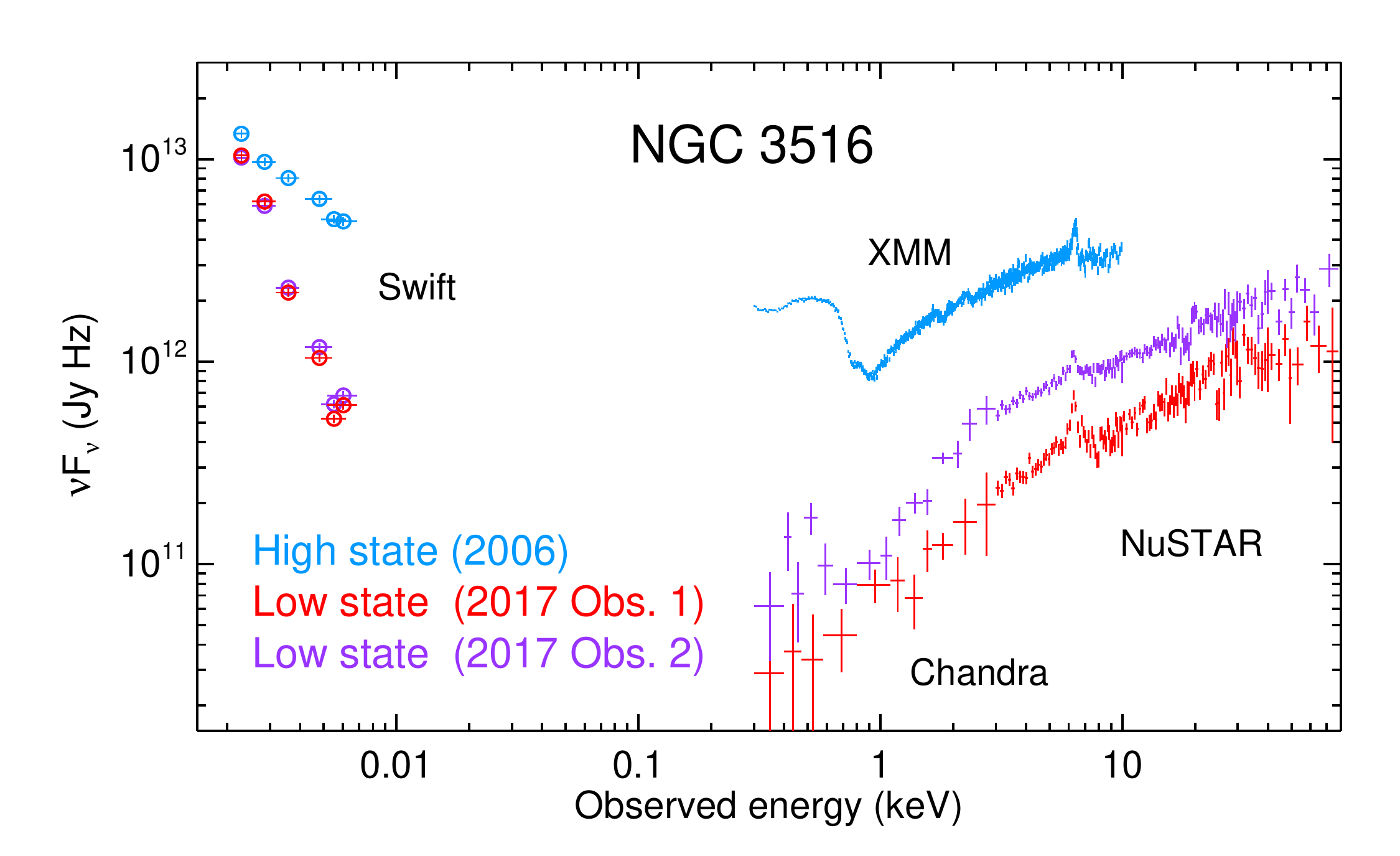}}
\caption{Optical/UV and X-ray spectra of \ngc taken during the high-flux state in 2006 and the low-flux state in 2017. The data are from \chandra/LETG (2017), \nustar (2017), \xmm/EPIC-pn (2006), and \swift/UVOT (2006 and 2017). The displayed data include the host galaxy emission and the effects of intrinsic and Galactic absorption and reddening. The displayed X-ray spectra have been binned for clarity of presentation. The best-fit model to these data is shown in Fig. \ref{fig_fit} and the derived intrinsic continuum model in Fig. \ref{fig_SED}. Transformation in shape and flux of the multi-wavelength spectrum between the two epochs is evident.
\label{fig_spec}}
\vspace{0.2cm}
\end{figure}

The observation logs of our \ngc data are provided in Table \ref{table_log}. We model two sets of joint \chandra/LETG and \nustar observations, which we refer to as `Obs. 1' and `Obs. 2'. The two observations were taken in December 2017, separated by about three weeks. We also include the associated contemporaneous \swift/UVOT data in our modeling of each observation. In Fig. \ref{fig_spec} the new \swift/UVOT, \chandra/LETG, and \nustar spectra (Obs. 1 and 2) taken during the changing-look low-flux state in December 2017 are compared with the \swift/UVOT and \xmm/EPIC-pn spectra taken at the high-flux state in 2006. The transformation in shape and flux of the optical-UV-X-ray spectrum between the two epochs is striking. By modeling the new 2017 spectra, and comparing the results with those from the archival 2006 observations, we investigate the effects of the changing-look event on the intrinsic continuum and the ionized outflows in \ngc.

For scheduling reasons each \chandra and \nustar observation was split into three separate exposures spanning a few days (Table \ref{table_log}). For our spectral modeling we stack the individual exposures of each observation to improve the signal-to-noise ratio. The variability between the individual exposures is not too much to prevent stacking of the spectra. All X-ray spectra were optimally binned according to \citet{Kaas16} for fitting in \spex. For each 2017 observation we jointly fitted the spectra from \chandra/LETG (0.3--3 keV), \nustar (3--78 keV), and the \swift/UVOT optical/UV filters. 

In \citet{Meh10} we studied the warm-absorber in \ngc using \xmm RGS and EPIC-pn spectra taken in 2006 during the normal high-flux state. To compare the SED models of the low-flux and high-flux states, we also fit one of the 2006 \xmm/EPIC-pn spectra (Obs. ID: 0401210501, 68~ks) over 0.3--10 keV. The data reduction of this EPIC-pn spectrum is described in \citet{Meh10}. As the Optical Monitor (OM) data of the 2006 \xmm observations were all taken in just one filter ($U$), for the purpose of deriving the optical/UV part of the SED we make use of the \swift/UVOT exposures that were taken in six optical and UV filters in 2006 ($V$, $B$, $U$, $UVW1$, $UVM2$, and $UVW2$), and fit them together with the 2006 EPIC-pn spectrum. The reduction and processing of the \chandra/LETG, \nustar, and \swift data are described as follows in Sects. \ref{sect_chandra}, \ref{sect_nustar}, and \ref{sect_swift}, respectively.

%============================
% TABLE: data log
%
\begin{deluxetable}{c c c c}
\tablecaption{Log of our \ngc observations taken in 2017. The `Obs. 1' and `Obs. 2' refer to two sets of spectra, separated by about three weeks, which we model in this paper. 
\label{table_log}}
\tablewidth{0pt}
\tablehead{
\colhead{} & \colhead{} & \colhead{Obs. date} & \colhead{Length} \\
\colhead{Observatory} & \colhead{Obs. ID} & \colhead{yyyy-mm-dd} & \colhead{(ks)}
}
\startdata
\chandra/LETG & 20877 (Obs. 1)	& 2017-12-05 & 24.9 \\
\chandra/LETG & 19519 (Obs. 1)	& 2017-12-07 & 35.8 \\
\chandra/LETG & 20878 (Obs. 1)	& 2017-12-11 & 21.0 \\
\chandra/LETG & 20904 (Obs. 2)	& 2017-12-26 & 35.8 \\
\chandra/LETG & 19520 (Obs. 2) & 2017-12-29 & 19.5 \\
\chandra/LETG & 20905 (Obs. 2)	& 2017-12-30 & 27.5 \\
\hline
\nustar & 60302016002 (Obs. 1)	& 2017-12-05 & 28.8 \\
\nustar & 60302016004 (Obs. 1)	& 2017-12-07 & 32.7 \\
\nustar & 60302016006 (Obs. 1)	& 2017-12-11 & 33.0 \\
\nustar & 60302016008 (Obs. 2)	& 2017-12-26 & 22.3 \\
\nustar & 60302016010 (Obs. 2)	& 2017-12-28 & 34.1 \\
\nustar & 60302016012	(Obs. 2) & 2017-12-30 & 33.3 \\
\hline
\swift & 00035462010	& 2017-12-04 & 1.5 \\
\swift & 00035462012 	& 2017-12-13 & 0.8 \\
\swift & 00035462013	& 2017-12-16 & 1.1 \\
\swift & 00035462014 	& 2017-12-30 & 1.2 \\
\enddata
\tablecomments{In addition to our 2017 \swift observations shown above, we make use of other \swift data, spanning from 2006 to 2021, to help characterize the long-term X-ray and optical/UV variability of \ngc. We also fit an archival 2006 \xmm/EPIC-pn spectrum (Obs. ID: 0401210501, 68~ks), taken from \citet{Meh10}, where we studied the warm-absorber of \ngc.}
\end{deluxetable}
%============================

%============================
% FIG: Swift, LETG, NUSTAR spectra
%
%============================

%%%%%%%%%%%%%%%%%%%%%%%%%%%%%%%%%%%%%%%%%%%%%%%%%%%%%%%%%%%%%%%%%%%%%%%%%%%%%%%%%%%%%%%%%%%%%%%%%%%%%%%
\subsection{Chandra/LETG data}
\label{sect_chandra}

Our \chandra observations of \ngc were taken with the LETG \citep{Brink00} grating and the HRC-S camera. The data were reduced using the Chandra Interactive Analysis of Observations (CIAO, \citealt{Frus06}) v4.9 software and the HEASARC Calibration Database (CALDB) v4.7.3. The {\tt chandra\_repro} script of CIAO and its associated tools were used for the reduction of the data and production of the final grating products (PHA2 spectra, RMF and ARF response matrices). The +/- first-order spectra and their response matrices were combined using the CIAO {\tt combine\_grating\_spectra} script. 

%%%%%%%%%%%%%%%%%%%%%%%%%%%%%%%%%%%%%%%%%%%%%%%%%%%%%%%%%%%%%%%%%%%%%%%%%%%%%%%%%%%%%%%%%%%%%%%%%%%%%%%
\subsection{NuSTAR data}
\label{sect_nustar}

The \nustar \citep{Harr13} observations were reduced using the NuSTAR Data Analysis Software (NUSTARDAS v.1.8.0) and CALDB calibration files of HEASoft v6.22. The data were processed with the standard pipeline script {\tt nupipeline} to produce level 1 calibrated and level 2 cleaned event files. The data from the South Atlantic Anomaly passages were filtered out and event files were cleaned with the standard depth correction. The source was extracted from a circular region (radius $= 75''$), with the background extracted from a source-free area of equal size on the same detector. The {\tt nuproducts} script was run to create level 3 products (spectra, ARF and RMF response files) for each of the two hard X-ray telescope modules (FPMA and FPMB) onboard \nustar. For each \nustar observation spectra and corresponding response matrices of the two telescopes were combined using the {\tt mathpha}, {\tt addrmf}, and {\tt addarf} tools of the HEASoft package.

%%%%%%%%%%%%%%%%%%%%%%%%%%%%%%%%%%%%%%%%%%%%%%%%%%%%%%%%%%%%%%%%%%%%%%%%%%%%%%%%%%%%%%%%%%%%%%%%%%%%%%%
\subsection{Swift XRT and UVOT data}
\label{sect_swift}

For the \swift \citep{Gehr04} X-ray Telescope (XRT, \citealt{Burr05}) data reduction, we used the procedure detailed in \citet{Evan07,Eva09}, which is an enhanced version of the standard \swift processing pipeline including some modifications. This tool is made available online on the UK Swift Science Data Centre (UKSSDC). The XRT instrument was operated in the Photon Counting (PC) mode. The default grades of 0--12 in the PC mode were used for event selection. The data were corrected for bad pixels and effects of vignetting and PSF to produce cleaned event files. The optimum extraction radius for data products depends on the count rate as reported in \citet{Evan07,Eva09}, which for \ngc mostly corresponds to an extraction radius of $70.8''$. The XRT lightcurves at soft (0.3--1.5 keV) and hard (1.5--10 keV) X-ray energies were constructed from each \swift snapshot, as described in \citet{Evan07}.

The \swift UVOT \citep{Romi05} data from Image-mode operations were taken with the six primary photometric filters of $V$, $B$, $U$, $UVW1$, $UVM2$ and $UVW2$. The $\mathtt{uvotsource}$ tool was used to perform aperture photometry using a circular aperture diameter of 10$\arcsec$. The standard instrumental corrections and calibrations according to \citet{Poo08} were applied. For the purpose of spectral fitting with {\tt SPEX}, the count rate and the corresponding response matrix file for each filter were created.
%%%%%%%%%%%%%%%%%%%%%%%%%%%%%%%%%%%%%%%%%%%%%%%%%%%%%%%%%%%%%%%%%%%%%%%%%%%%%%%%%%%%%%%%%%%%%%%%%%%%%%%
%%%%%%%%%%%%%%%%%%%%%%%%%%%%%%%%%%%%%%%%%%%%%%%%%%%%%%%%%%%%%%%%%%%%%%%%%%%%%%%%%%%%%%%%%%%%%%%%%%%%%%%
%%%%%%%%%%%%%%%%%%%%%%%%%%%%%%%%%%%%%%%%%%%%%%%%%%%%%%%%%%%%%%%%%%%%%%%%%%%%%%%%%%%%%%%%%%%%%%%%%%%%%%%
\section{Spectral modeling} 
\label{sect_model}

Here we present our modeling of the spectral components that form the observed SED in \ngc. This consists of simultaneously modeling the continuum components (described in Section \ref{sect_cont}) and the X-ray absorption components (described in Section \ref{sect_abs}). We derive the SED and the intrinsic X-ray absorption for Obs. 1 and Obs. 2, using \chandra/LETG, \nustar, and \swift/UVOT data, taken in December 2017 during the low-flux state of \ngc. For comparison with the 2017 model, we also model the archival \xmm/EPIC-pn and \swift/UVOT data taken in 2006 (described in Section \ref{sect_data}) to obtain the SED in the normal high-flux state of \ngc.   

%%%%%%%%%%%%%%%%%%%%%%%%%%%%%%%%%%%%%%%%%%%%%%%%%%%%%%%%%%%%%%%%%%%%%%%%%%%%%%%%%%%%%%%%%%%%%%%%%%%%%%%
\subsection{Broadband continuum modeling}
\label{sect_cont}

\begin{figure*}
\centering
\begin{minipage}[c]{0.49\linewidth}
\hspace{-0.33cm}\resizebox{1.054\hsize}{!}{\includegraphics[angle=0]{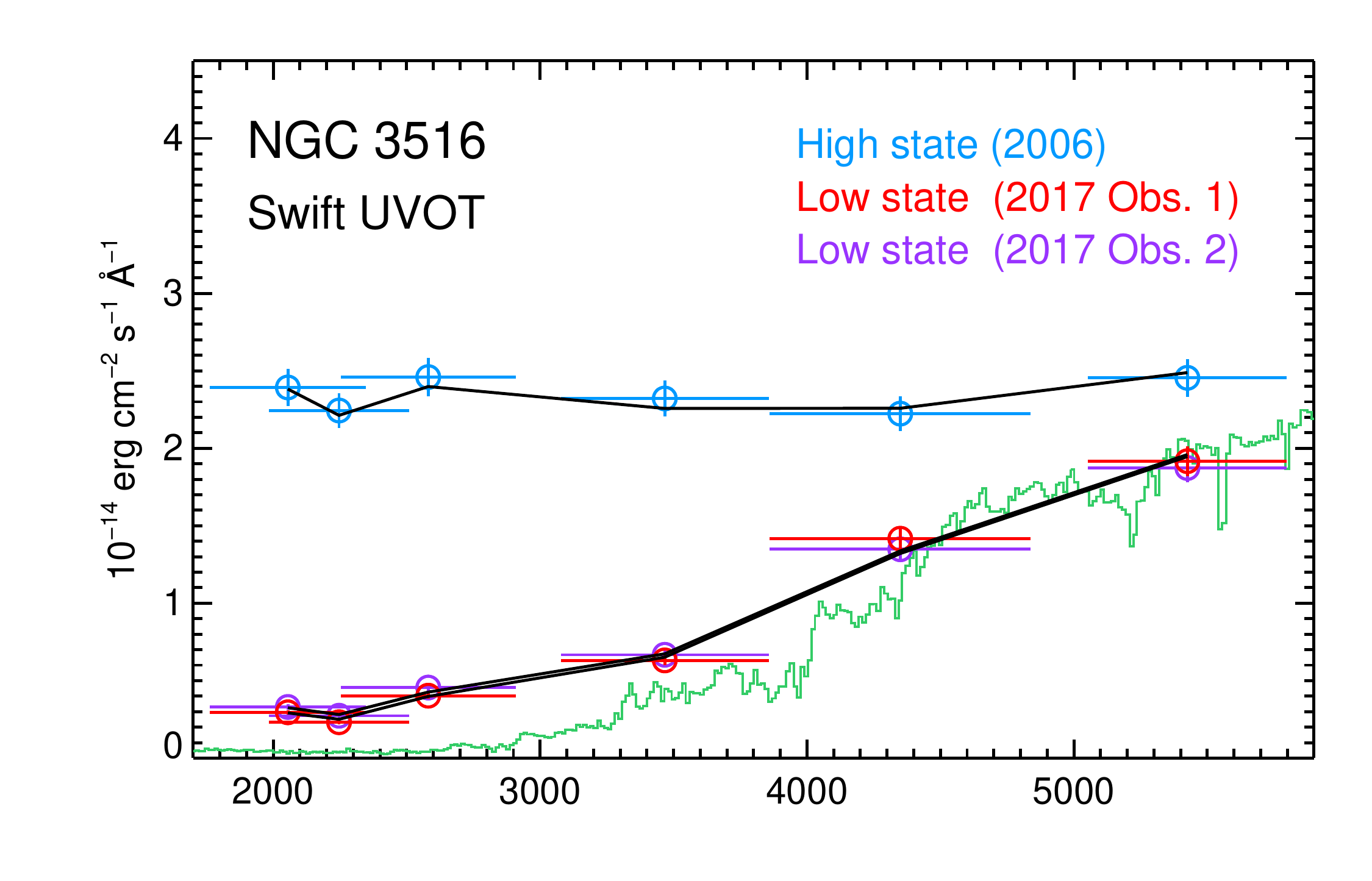}}

\vspace{-0.89cm}
\hspace{-0.33cm}\resizebox{1.054\hsize}{!}{\includegraphics[angle=0]{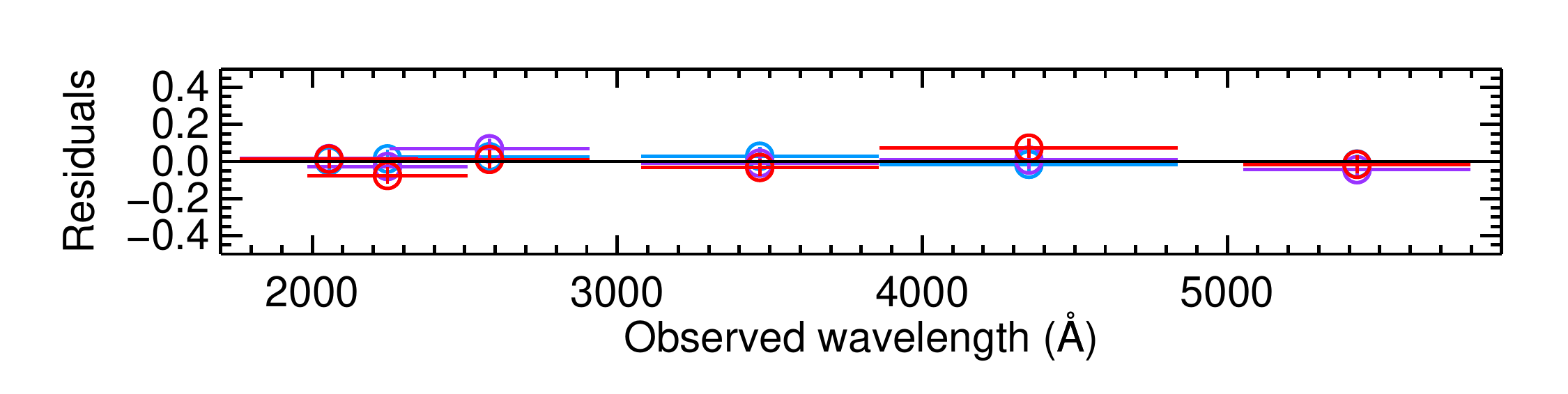}}
\end{minipage}
\begin{minipage}[c]{0.49\linewidth}

\vspace{0.27cm}
\resizebox{\hsize}{!}{\includegraphics[angle=270]{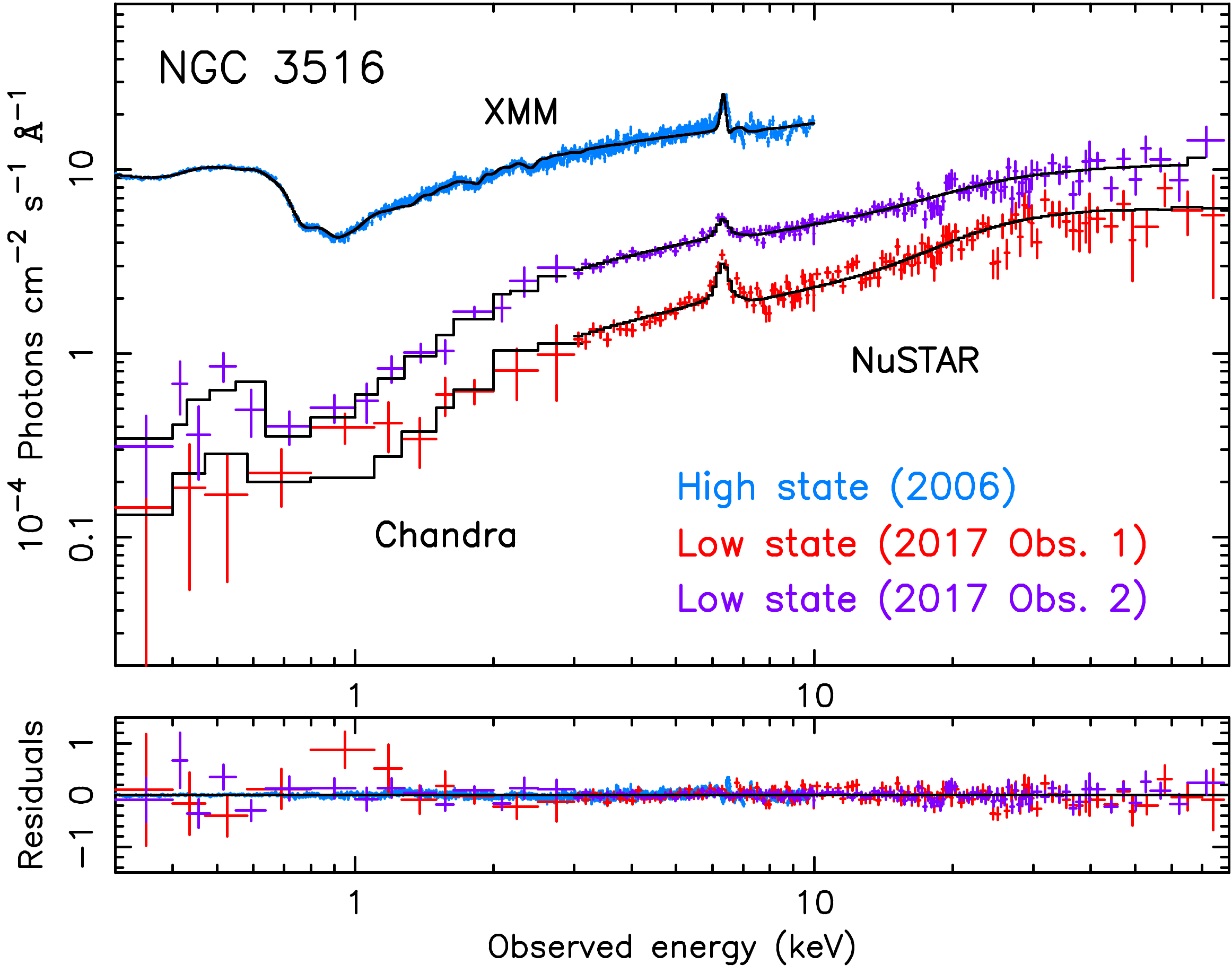}}
\end{minipage}
\caption{Best-fit \spex model to the \swift/UVOT, \chandra/LETG, and \nustar spectra of Obs. 1 and Obs. 2 taken during the low-flux state in December 2017, compared with the best-fit model to the \swift/UVOT and \xmm/EPIC-pn data taken in the high-flux state in 2006. The optical/UV data are shown in the {\it left panels} and the X-ray data in the {\it right panels}. The best-fit model for each dataset is shown as a black line. Residuals of the fit, defined as (data$-$model)/model, are displayed in the {\it bottom panels}. The model for the host galaxy stellar emission is shown in green in the {\it top left panel}. The displayed X-ray spectra have been binned for clarity of presentation. The intrinsic broadband continuum models corresponding to these best-fits are shown in Fig. \ref{fig_SED}.
\label{fig_fit}}
\vspace{0.5cm}
\end{figure*}

The broadband continuum model that we apply to fit the \ngc data (Fig. \ref{fig_fit}) consists of three continuum components in \spex: (1) a {\tt comt} component that models the optical/UV thermal continuum from the accretion disk and the associated `soft X-ray excess' by warm Comptonization; (2) a {\tt pow} component that models the primary X-ray power-law; (3) a {\tt refl} neutral reflection component that produces the \FeKa line and the Compton hump at hard X-rays. Such modeling of the broadband continuum has previously been carried out for other similar AGN, such as the archetypal Seyfert-1 galaxy NGC~5548 \citet{Meh15a}. 

Our previous \xmm study of \ngc \citep{Meh10} shows a clear presence of a `soft X-ray excess' component in the normal high-flux state of \ngc in 2006. Here we model the soft excess in \ngc with warm Comptonization, which multi-wavelength studies of other similar Seyfert-1 AGN have found to be a viable explanation (see e.g. \citealt{Meh11,Done12,Meh15a,Porq18,Kubo18,Petr18,Petr20}). In this explanation of the soft excess, the seed disk photons are up-scattered in a warm, optically thick, corona to produce the extreme UV (EUV) continuum and the soft X-ray excess as its tail at higher energies. We note that this is one plausible explanation proposed in the literature for the origin of the soft X-ray excess in AGN (relativistic ionized reflection is another explanation, \citealt{Cru06}), yet it is sufficient for our purpose of comparing the high-flux and low-flux states of \ngc SEDs. The parameters of the {\tt comt} model are its normalization, seed photon temperature $T_{\rm seed}$, electron temperature $T_{\rm e}$, and optical depth $\tau$ of the up-scattering corona. To limit the number of free parameters while still providing a good fit, some of the {\tt comt} parameters of different observations are coupled together as shown in Table \ref{table_fit}, where the best-fit parameters of the continuum components are given.

The X-ray power-law ({\tt pow}) represents Compton up-scattering of the seed disk photons in an optically-thin and hot corona. For each observation we fit the normalization and the photon index $\Gamma$ of the {\tt pow} model. The intrinsic power-law continuum then undergoes reprocessing, modeled by the {\tt refl} reflection model in \spex, which computes the \FeKa line according to \citet{Zyck94}, and the Compton-reflected continuum according to \citet{Magd95}, as described in \citet{Zyck99}. For the 2006 observation the normalization and the photon index $\Gamma$ of the illuminating power-law for {\tt refl} are coupled to those of the {\tt pow} component. For the 2017 observations (i.e. Obs. 1 and 2) the normalization and $\Gamma$ of {\tt refl} are coupled to the average of the {\tt pow} model for the two observations, representing a time-averaged illuminating power-law that is reflected in the 2017 epoch.

The high-energy exponential cut-off ($E_{\rm cut}$) of {\tt pow} and the incident power-law for {\tt refl} were coupled together and fitted as one parameter for all observations. We find the best-fit model to the \nustar spectra favors $E_{\rm cut} > 1$~MeV, which minimizes excess fit residuals at the higher-energy part of the \nustar spectra. We thus fixed $E_{\rm cut}$ to 1 MeV in our SED modeling. A low-energy exponential cut-off was also applied to the power-law continuum at the far-UV (13.6 eV) to prevent it exceeding the energy of the seed photons from the disk. The ionization parameter of {\tt refl} was set to zero to produce a neutral reflection component, which is consistent with the observed neutral \FeKa line in \ngc \citep{Meh10}. In our modeling we fitted the reflection fraction parameter of the {\tt refl} model for each observation.

In our modeling of the optical/UV data we took into account the host galaxy starlight emission and the contribution of AGN emission lines in the \swift/UVOT filters. We used the galactic bulge template model of \citet{Kin96}, and the \ngc host galaxy flux measured by \citet{Ben13} from HST imaging, to calculate the corresponding stellar emission model for the 10\arcsec circular aperture of UVOT. This model is shown in Fig. \ref{fig_fit} (top left panel). To correct for the contribution of the AGN emission lines to the UVOT photometric filters (2 to 5\%), we used the template model derived in \citet{Meh15a} for the archetypal Seyfert-1 galaxy NGC~5548, and normalized this model to the H$\beta$ flux of \ngc. The H$\beta$ flux was taken from the long-term optical monitoring study of \ngc by \citet{Shap19}, which reports that during the 2006 high-flux state the H$\beta$ flux is about ${9 \times 10^{-13}}$~\ergflux and in the 2017 low-flux state is about ${1 \times 10^{-13}}$~\ergflux. We correct for the foreground Milky Way extinction in our line of sight to \ngc, ${E(B - V) = 0.037}$ \citep{Schl11}, using the {\tt ebv} model in \spex, which uses the extinction curve of \citet{Car89}, including the update for near-UV given by \citet{ODo94}. The ratio of total to selective extinction ${R_V = A_V/E(B-V)}$ was fixed to 3.1.

Figure \ref{fig_fit} shows our best-fit model to the 2006 (\swift/UVOT and \xmm/EPIC-pn) and the 2017 Obs. 1 and Obs. 2 (\swift/UVOT, \chandra/LETG, and \nustar) spectra. The modeling takes into account the X-ray absorption described below in Section \ref{sect_abs}. The best-fit parameters of the {\tt comt}, {\tt pow}, and {\tt refl} components for each observation are given in Table \ref{table_fit}. The corresponding intrinsic luminosities of the continuum components for each observation are shown in Table \ref{table_lum}. Figure \ref{fig_SED} displays the intrinsic continuum SED model derived from the 2006 and the 2017 Obs. 1 and 2 data, showing how the individual continuum components change between the observations. All continuum components of the \ngc SED ({\tt comt}, {\tt pow}, and {\tt refl}) have become fainter in the low-flux changing-look state seen in 2017.

%============================
% FIG: Best-fit
%
%============================

%============================
% TABLE: best-fit parameters
%
\begin{deluxetable}{c | c}
\tablecaption{Best-fit parameters of the broadband continuum model components for \ngc, derived from modeling the normal high-flux state observation taken in 2006, and the low-flux state Obs. 1 and 2 taken in 2017.
\label{table_fit}}
\tablewidth{0pt}
\tablehead{
Parameter & \colhead{Value}
}
\startdata
\multicolumn{2}{c}{Disk component: optical-UV and the soft X-ray excess ({\tt comt}):} 						\\
Normalization	&		${2.9 \pm 0.1}$ (2006) \\
             	&		${0.3 \pm 0.1}$ (Obs. 1) \\
             	&		${0.4 \pm 0.1}$ (Obs. 2) \\
$T_{\rm seed}$ (eV) &	${0.70 \pm 0.02}$ (2006) \\
                    & ${0.46 \pm 0.08}$	(Obs. 1 and 2, coupled) \\
$T_{\rm e}$ (keV) &		${0.13 \pm 0.01}$ (2006) \\
                    & ${0.13 \pm 0.01}$	(Obs. 1 and 2, coupled) \\
                    Optical depth $\tau$ &	${28 \pm 3}$ (all observations, coupled) \\
\hline
\multicolumn{2}{c}{Primary X-ray power-law component ({\tt pow}):} 						\\
Normalization			& ${27.3 \pm 0.1}$ (2006) \\
            			& ${2.6 \pm 0.2}$ (Obs. 1) \\
            			& ${5.7 \pm 0.4}$ (Obs. 2) \\
Photon index $\Gamma$		& ${1.87 \pm 0.02}$	(2006)		\\
                    		& ${1.75 \pm 0.02}$	(Obs. 1)		\\
                    		& ${1.73 \pm 0.02}$	(Obs. 2)		\\
\hline
\multicolumn{2}{c}{X-ray reflection component ({\tt refl}):} 						\\
Reflection fraction			& ${0.53 \pm 0.03}$ (2006)					\\
                    		  & ${0.61 \pm 0.03}$	(Obs. 1)		\\
                    		  & ${0.61 \pm 0.03}$	(Obs. 2)		\\
\hline
\multicolumn{2}{c}{C-stat\,/ expected C-stat = 1758\,/\,1100 (2006)}  \\
\multicolumn{2}{c}{C-stat\,/ expected C-stat = 2218\,/\,1896 (Obs. 1)}  \\
\multicolumn{2}{c}{C-stat\,/ expected C-stat = 2184\,/\,1908 (Obs. 2)}  \\
\enddata
\tablecomments{The normalization of the Comptonized disk component ({\tt comt}) is in units of $10^{55}$ photons~s$^{-1}$~keV$^{-1}$. The power-law normalization of the {\tt pow} and {\tt refl} components is in units of $10^{50}$ photons~s$^{-1}$~keV$^{-1}$ at 1 keV. For the 2006 observation the normalization and the photon index $\Gamma$ of the incident power-law for {\tt refl} are coupled to those of the primary power-law {\tt pow}. For Obs. 1 and 2 the normalization and $\Gamma$ of {\tt refl} were coupled to the average of the {\tt pow} model for the two observations, representing a time-averaged illuminating power-law for reflection in the 2017 epoch. The high-energy exponential cut-off of the power-law for both {\tt pow} and {\tt refl} is fixed to 1~MeV as described in Section \ref{sect_cont}. The intrinsic luminosities of the continuum components for each observation are provided in Table \ref{table_lum}.}
\end{deluxetable}
%============================

%============================
% TABLE: Luminosities
%
\begin{deluxetable}{c | c c c}
\tablecaption{Intrinsic luminosities of \ngc for the normal high-flux state in 2006, and the changing-look low-flux state in 2017 Obs. 1 and 2. The intrinsic luminosities are derived from our broadband continuum modeling described in Section \ref{sect_cont}, using the Comptonized disk ({\tt comt}), power-law ({\tt pow}), and the reflection ({\tt refl}) components, shown in Fig. \ref{fig_SED} and Table \ref{table_fit}.
\label{table_lum}}
\tablewidth{0pt}
\tablehead{
 & \colhead{2006} & \colhead{2017 Obs. 1} & \colhead{2017 Obs. 2} \\
Luminosity & \colhead{($10^{43}$~\ergs)} & \colhead{($10^{43}$~\ergs)} & \colhead{($10^{43}$~\ergs)}
}
\startdata
$L_{\tt comt}$	&	5.5 & 0.5 & 0.6 \\
$L_{\tt pow}$	  &	5.7 & 0.8 & 1.9 \\
$L_{\tt refl}$	&	0.8 & 0.2 & 0.2 \\
\hline
$L_{\rm opt}$	  &	0.5 & 0.04 & 0.06 \\
$L_{\rm UV}$	  &	1.7 & 0.1 & 0.2 \\
$L_{\rm EUV}$	  &	2.9 & 0.2 & 0.3 \\
$L_{\rm soft}$	& 1.3 & 0.1 & 0.2 \\
$L_{\rm hard}$	&	1.0 & 0.1 & 0.3 \\
$L_{\rm ion}$	  &	6.4 & 0.6 & 1.0 \\
$L_{\rm bol}$	  &	12.0 & 1.5 & 2.7 \\
\enddata
\tablecomments{The luminosities $L_{\tt comt}$, $L_{\tt pow}$, and $L_{\tt refl}$ are calculated over the entire energy band ($10^{-6}$--$10^{6}$ keV). The intrinsic luminosity in the optical band ($L_{\rm opt}$) is calculated over 4000--7000 \AA, UV ($L_{\rm UV}$) over 1000--4000 \AA, EUV ($L_{\rm EUV}$) over 100--1000 \AA, soft X-ray ($L_{\rm soft}$) over 0.3--1.5 keV, and hard X-ray ($L_{\rm hard}$) over 1.5--10 keV. The ionizing luminosity $L_{\rm ion}$ is calculated over 1--1000 Ryd, which is used in the definition of the ionization parameter $\xi$. The bolometric luminosity $L_{\rm bol}$ is the total intrinsic luminosity of the broadband continuum over $10^{-6}$--$10^{6}$ keV, calculated as the sum of the intrinsic luminosities of the continuum components $L_{\tt comt}$, $L_{\tt pow}$, and $L_{\tt refl}$.}
\end{deluxetable}
%============================

%============================
% TABLE: Warm-absorber parameters
%
\begin{deluxetable}{c | c c c}
\tablecaption{Column density \NH and ionization parameter $\xi$ of the \ngc warm-absorber components in the normal 2006 high-flux state, and the changing-look 2017 (Obs. 1 and 2) low-flux state. Because of the dimming of the ionizing continuum in 2017 (Fig. \ref{fig_SED}) the warm-absorber becomes `de-ionized' as described in Section \ref{sect_abs}. This results in additional X-ray absorption by the warm-absorber as shown in Fig. \ref{fig_trans}.
\label{table_WA}}
\tablewidth{0pt}
\tablehead{
Parameter & \colhead{2006} & \colhead{2017 Obs. 1} & \colhead{2017 Obs. 2}
}
\startdata
Comp 1:	&	& &  \\
\NH	        &	$1.0 \pm 0.2$ & 1.0 (fixed) & 1.0 (fixed) \\
$\log \xi$	&	$2.8 \pm 0.1$ & 1.8 (de-ionized) & 2.0 (de-ionized) \\
\hline
Comp 2:	&	& &  \\
\NH	        &	$1.7 \pm 0.2$ & 1.7 (fixed) & 1.7 (fixed) \\
$\log \xi$	&	$2.2 \pm 0.1$ & 1.1 (de-ionized) & 1.3 (de-ionized) \\
\hline
Comp 3:	&	& &  \\
\NH	         & $0.36 \pm 0.01$ & 0.36 (fixed) & 0.36 (fixed) \\
$\log \xi$	 & $0.8 \pm 0.1$ & $-0.5$ (de-ionized) & $-0.2$ (de-ionized) \\
\enddata
\tablecomments{Column density \NH is given in $10^{22}$~\cm. The model is based on the warm-absorber study of \citet{Meh10} using the 2006 \xmm observation of \ngc as described in Section \ref{sect_abs}.}
\vspace{-0.3cm}
\end{deluxetable}
%============================

%%%%%%%%%%%%%%%%%%%%%%%%%%%%%%%%%%%%%%%%%%%%%%%%%%%%%%%%%%%%%%%%%%%%%%%%%%%%%%%%%%%%%%%%%%%%%%%%%%%%%%%
\subsection{X-ray absorption and photoionization modeling}
\label{sect_abs}
In our X-ray absorption modeling we first take into account the continuum and line absorption by the diffuse interstellar medium (ISM) in the Milky Way. This is done using the {\tt hot} model in \spex \citep{dePl04,Stee05}. This model calculates the transmission of a plasma in collisional ionization equilibrium at a given temperature, which for neutral ISM is
set to the minimum temperature of the model at 0.008~eV. The total Galactic \NH column density was fixed to $4.04 \times 10^{20}$~\cm \citep{Will13}, which is the sum of the atomic and molecular hydrogen components of the ISM in our line of sight to \ngc.

Our modeling of the ionized warm-absorber in \ngc is carried out using the \pion model \citep{Meh16b,Mill15} in \spex. This self-consistent model calculates both the photoionization equilibrium solution and the spectrum, using the SED model that is simultaneously fitted to the data in \spex. Previous spectroscopic studies of \ngc (e.g. \citealt{Tur08,Meh10,Holc12}) have shown the clear presence of intrinsic X-ray absorption by the warm-absorber in this AGN. In \citet{Meh10} we derived a model for the warm-absorber using \xmm (RGS and EPIC-pn) observations taken in 2006. During this epoch \ngc was in its normal high-flux state, and thus the features of the warm-absorber were detected with high signal-to-noise ratio. The modeling of \citet{Meh10} shows that the warm-absorber consists of three ionization components. 

Here we make use of the warm-absorber model derived by \citet{Meh10}. However, in \citet{Meh10} an older version of the \spex code was used (v2.01.02), and over the past decade there have been significant advances and changes in the atomic data and plasma models of the \spex code. Therefore, since here we are using the latest code, we re-fitted the 2006 spectrum. The column density \NH and ionization parameter $\xi$ (defined by \citealt{Kro81}) of the three warm-absorber components of \citet{Meh10} were re-fitted to take into account changes between the codes. The other warm-absorber parameters (velocities and covering fractions) were kept to those measured by \citet{Meh10}. In our modeling in this paper the abundances of the warm-absorber, and that of the Galactic absorption, are fixed to the proto-solar values of \citet{Lod09}. In our model set-up, the ionizing SED first passes through the highest ionization component of the warm-absorber (Comp 1) and last through the lowest one (Comp 3). The \NH and $\xi$ of the warm-absorber components for the 2006 observation are given in Table \ref{table_WA}.

%============================
% FIG: SED
%
\begin{figure*}
\centering
\resizebox{\hsize}{!}{\includegraphics[angle=0]{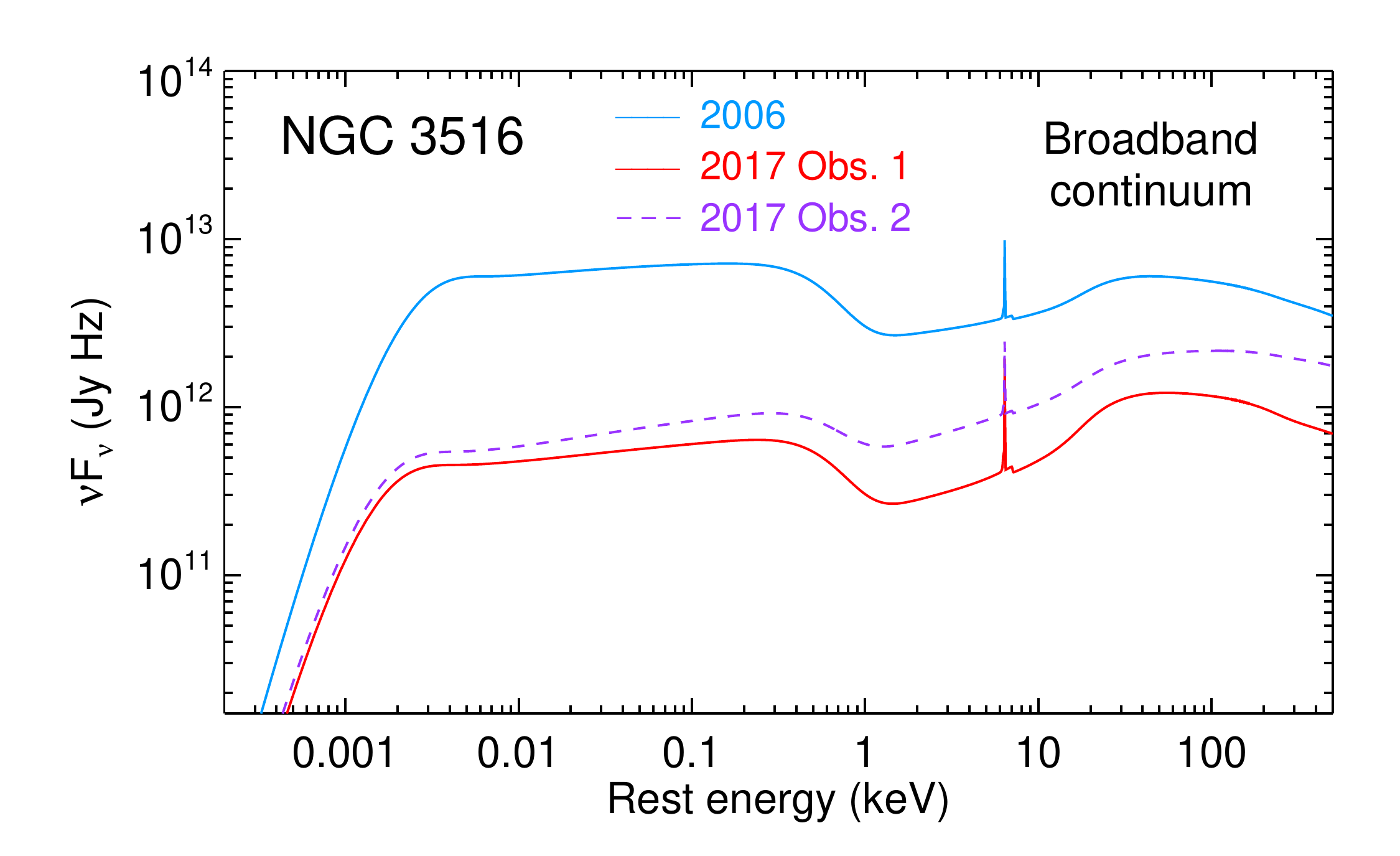}\includegraphics[angle=0]{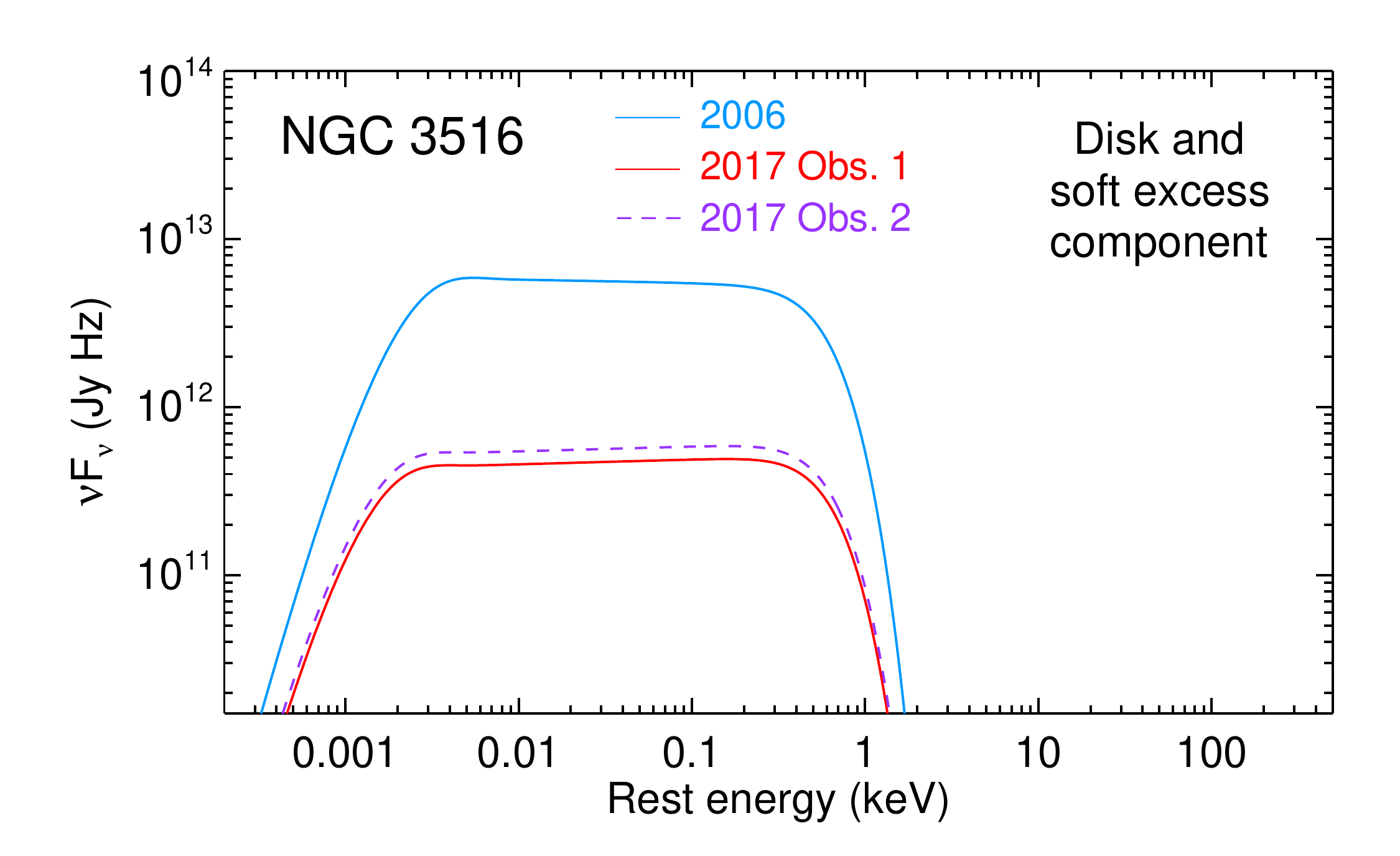}}\vspace{-0.4cm}
\resizebox{\hsize}{!}{\includegraphics[angle=0]{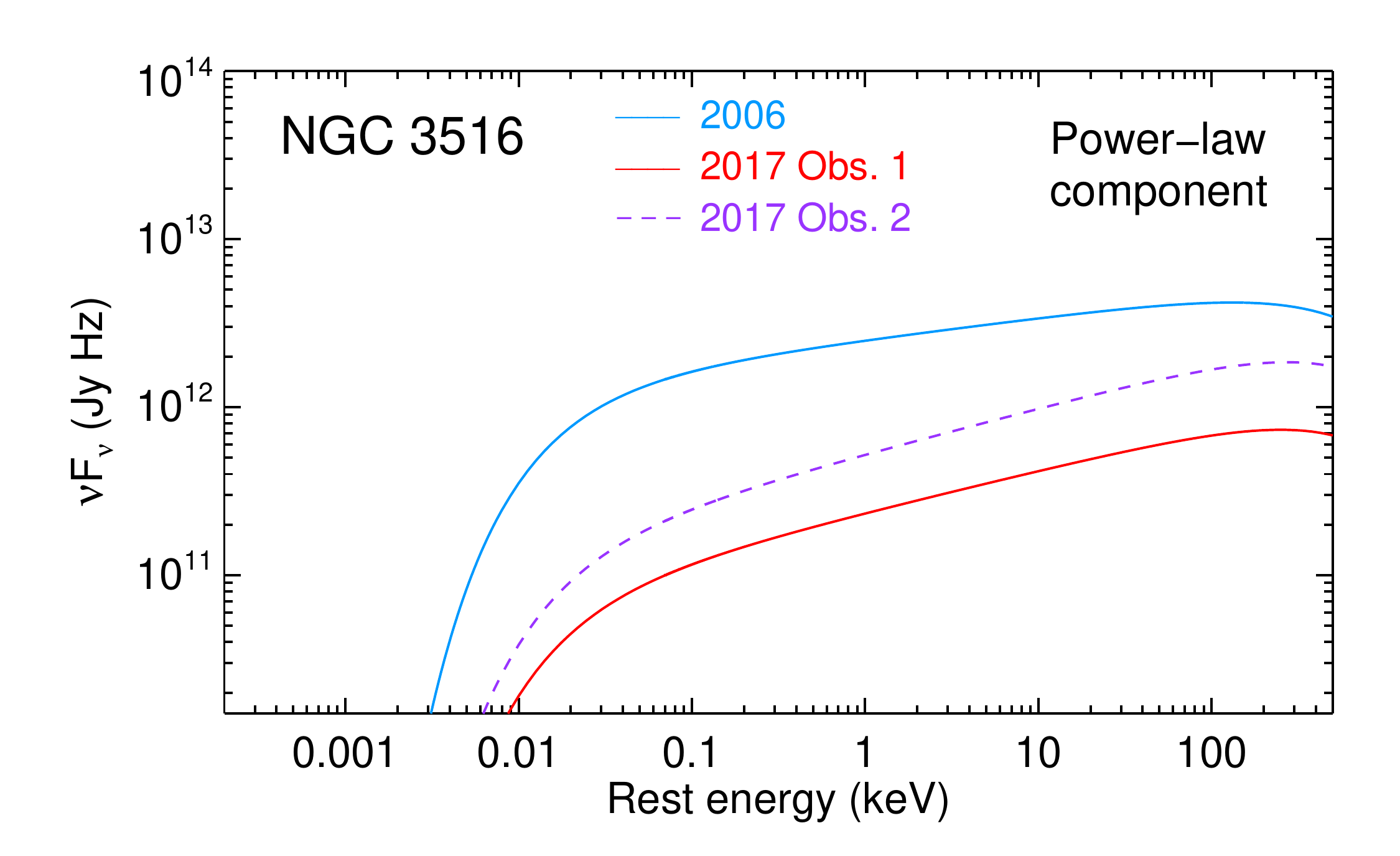}\includegraphics[angle=0]{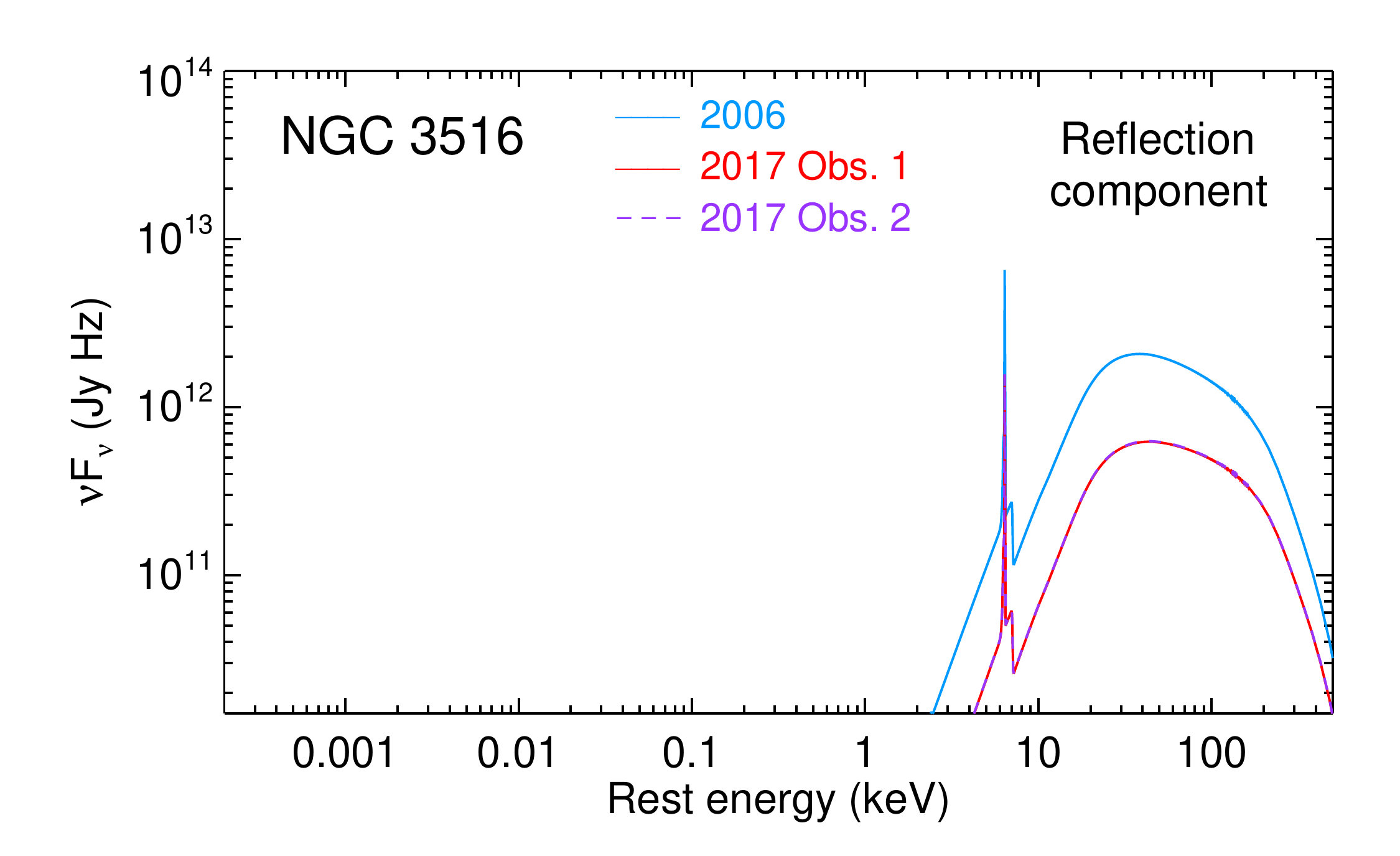}}
\caption{Derived intrinsic broadband SED model for the high-flux state (2006) and the low-flux state (2017) of \ngc. The total broadband continuum model is shown in the {\it top left panel}. The individual components of the SED model are displayed in the other panels: the disk and the soft X-ray excess component ({\tt comt}, {\it top right panel}), the X-ray power-law continuum component ({\tt pow}, {\it bottom left panel}), and the X-ray reflection component ({\tt refl}, {\it bottom right panel}). All continuum components of the SED have become fainter in the new changing-look state of \ngc in 2017. 
\label{fig_SED}}
\vspace{0.5cm}
\end{figure*}
%============================

In our modeling of the low-flux state 2017 Obs. 1 and 2 spectra, we adopt the above warm-absorber model, which is from a normal average-level \xmm observation taken in 2006 (Obs. ID: 0401210501). In the low-flux state of \ngc the X-ray continuum is strongly diminished (Figs. \ref{fig_spec}, \ref{fig_fit}, and \ref{fig_SED}). For this reason the individual absorption lines are not detected with sufficient signal-to-noise ratio to enable high-resolution X-ray spectral analysis. We therefore cannot independently model the warm-absorber to determine its properties, in particular the ionization parameter and the velocity. Thus, in our modeling of the 2017 data we keep the persistent warm-absorber parameters fixed to those obtained from 2006, while the ionization parameters $\xi$ of the components are self-consistently lowered in response to the intrinsic dimming of the SED. The total column density, covering fraction, and the velocity of the warm-absorbers in best-studied AGN (such as NGC~5548) are found to remain persistent over long times, while responding to the ionizing radiation (see e.g. \citealt{Arav15}). Similarly, in the case of \ngc, the warm-absorbers seen in X-rays in 1995 \citep{Kris96b} and 2006 \citep{Meh10} are consistent with each other. Therefore, for the 2017 observations the 2006 warm-absorber parameters are not re-fitted, but rather the ionization parameter of each component is automatically computed by {\tt pion} according to the low-flux SED of each observation in 2017 (Fig. \ref{fig_SED}). The lowered ionization parameters of the warm-absorber components for the 2017 observations are shown in Table \ref{table_WA}. Importantly, the warm-absorber in 2017 is `de-ionized', thus causing higher opacity and producing stronger absorption than in 2006, which can be seen in the X-ray transmission model shown in Fig. \ref{fig_trans}. Finally, we tested freeing $\xi$ of the warm-absorber components for the 2017 observations. This did not make a significant improvement to the best-fit model and the $\xi$ parameters remained consistent with the above `de-ionization' model.

Such de-ionization of the warm-absorber has been previously presented in NGC~5548 \citep{Kaas14,Arav15} and NGC~3783 \citep{Mehd17,Kris19}, in the context of shielding of the ionizing source by nuclear obscuring winds. However, here in the case of \ngc the de-ionization is instead caused by the dimming of the intrinsic SED as a consequence of the changing-look phenomenon. Interestingly, we find that as a result of the enhanced continuum absorption by the de-ionized warm-absorber, no additional absorption of the continuum is required to fit the 2017 X-ray spectra. Including a new absorption component (such as a partially-covering obscurer) does not significantly improve the fit to the data; $\Delta$C-stat improvement of only 10 by fitting three more free parameters which is not statistically reasonable for an additional component. We discuss the de-ionization of the warm-absorber in Section \ref{sect_outflows}.

Interestingly, we find that the flux of the X-ray emission lines in \ngc between the 2006 and 2017 observations has dropped significantly. The flux of the \FeKa line in the 2006 \xmm/EPIC-pn spectrum is ${7.6 \pm 0.6 \times 10^{-13}}$ \ergflux (this work and \citealt{Meh10}), while in the 2017 \nustar spectra it is: ${1.9 \pm 0.2 \times 10^{-13}}$ \ergflux. In \citet{Meh10} we reported a narrow \ion{O}{7} forbidden emission line at 22.1~\AA\ in the 2006 \xmm/RGS spectrum, as well as another emission feature likely corresponding to a broad and blue-shifted \ion{O}{8} Ly$\alpha$ line. The new \chandra/LETG spectra suggest that these emission lines have also become intrinsically fainter in 2017. The LETG spectra only allow us to constrain an upper limit on the flux of each line (${< 2 \times 10^{-14}}$ \ergflux for \ion{O}{7}, and ${< 3 \times 10^{-14}}$ \ergflux for \ion{O}{8}), which is lower than the flux previously measured by \citet{Meh10} in the 2006 high-flux state: ${5 \pm 1 \times 10^{-14}}$ \ergflux for \ion{O}{7}, and ${8 \pm 1 \times 10^{-14}}$ \ergflux for \ion{O}{8}.

%%%%%%%%%%%%%%%%%%%%%%%%%%%%%%%%%%%%%%%%%%%%%%%%%%%%%%%%%%%%%%%%%%%%%%%%%%%%%%%%%%%%%%%%%%%%%%%%%%%%%%%
%%%%%%%%%%%%%%%%%%%%%%%%%%%%%%%%%%%%%%%%%%%%%%%%%%%%%%%%%%%%%%%%%%%%%%%%%%%%%%%%%%%%%%%%%%%%%%%%%%%%%%%
%%%%%%%%%%%%%%%%%%%%%%%%%%%%%%%%%%%%%%%%%%%%%%%%%%%%%%%%%%%%%%%%%%%%%%%%%%%%%%%%%%%%%%%%%%%%%%%%%%%%%%%
\section{Discussion} 
\label{sect_discuss}

%%%%%%%%%%%%%%%%%%%%%%%%%%%%%%%%%%%%%%%%%%%%%%%%%%%%%%%%%%%%%%%%%%%%%%%%%%%%%%%%%%%%%%%%%%%%%%%%%%%%%%%
\subsection{Changing-look of the broadband SED in NGC 3516}
\label{sect_accretion}

\begin{figure}
\hspace{-0.40cm}\resizebox{1.075\hsize}{!}{\includegraphics[angle=0]{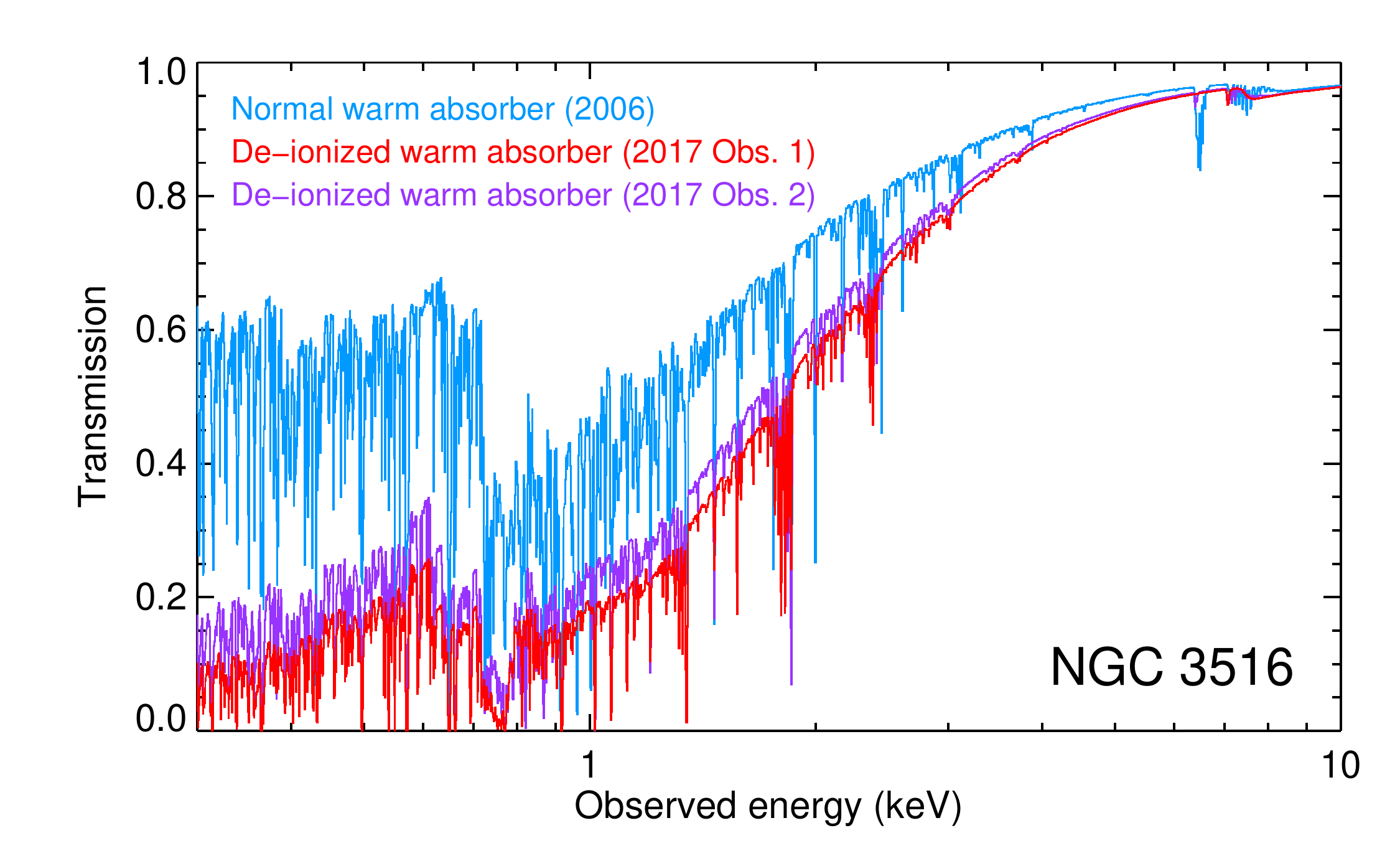}}
\caption{X-ray transmission (line and continuum absorption) of the \ngc warm-absorber model in the high-flux state (2006) and the low-flux state (2017). The dimming of the ionizing SED (Fig.~\ref{fig_SED}) in 2017 causes `de-ionization' of the warm-absorber, thus resulting in stronger X-ray continuum absorption (shown above) and hence a harder X-ray spectrum (Fig. \ref{fig_spec}). The transmission variability between Obs. 1 and 2, taken three weeks apart in 2017, is caused by variation in the ionization parameter $\xi$ of the warm-absorber components in response to the intrinsic continuum variability. This warm-absorber transmission variability results in additional X-ray hardness variability (Fig. \ref{fig_lc}).
\label{fig_trans}}
\vspace{0.3cm}
\end{figure}

Ionized gas in AGN is widely-accepted to be photoionized by the SED of the central ionizing source. Therefore, if this photoionizing SED changes it is prudent to first consider the unavoidable effect of this change on the already-present ionized gas in an AGN, before assuming new additional gas has appeared. The broadband (optical-UV-X-ray) continuum modeling presented in this paper shows that the ionizing SED in \ngc has changed significantly (Figs. \ref{fig_fit} and \ref{fig_SED}). The natural consequence of this SED change on the ionized warm-absorber gas is enhanced X-ray absorption by the warm-absorber according to photoionization computations (Fig. \ref{fig_trans}). The corresponding absorption model (i.e. predicted from photoionization modeling) matches well the spectrum of \ngc (Fig. \ref{fig_fit}), without needing to include any new absorbing gas in the fit.

From only considering the goodness-of-fit in the X-ray band, it is not practical to conclusively discern between the changing-continuum and the obscuration models. The obscuring gas in principle can have multiple partially-covering components with different ionization parameters and column densities (like found in NGC~5548 by \citealt{Kaas14}), thus by fitting these additional free parameters similarly good fits can be achieved. The changing-continuum model, on the other hand, does not need additional free parameters to make a good fit, but rather the absorption is self-consistently predicted according to realistic photoionization modeling. Furthermore, because in \ngc the flux level in the soft X-ray band is too low for high-resolution X-ray spectroscopy of absorption lines, one cannot analyze changes in the absorption lines, which is needed for a proper X-ray differentiation between the two models.

%============================
% FIG: Transmission spectrum
%
%============================

The X-ray obscuration scenario is limited in scope to the X-ray band and cannot explain the observed dimming of the optical/UV band at energies below the Lyman limit (Figs. \ref{fig_fit}, left panels). Nor does it explain the dimming of the hard X-ray continuum at $> 3$~keV as seen in our \nustar spectra (Fig. \ref{fig_fit}, right panels). On the other hand, the changing-continuum model provides a plausible multi-wavelength solution that can explain all the observed data from optical/UV to X-rays. Importantly, \ngc displays key differences from the established obscuration-event AGN, such as NGC~5548 \citep{Kaas14} and NGC~3783 \citep{Mehd17}, which also favor the changing-continuum explanation in \ngc as described in the following.

Dramatic spectral changes, revealed by changing-look AGN, provide us with useful probes for studying the nature of accretion-powered radiation and outflows in AGN. An important characteristic of the changing-look event in \ngc is the dimming of the intrinsic optical/UV continuum. This is in stark contrast to the transient obscuration events found in the Seyfert-1 galaxies \object{Mrk 335} \citep{Long13}, \object{NGC 5548} \citep{Kaas14}, \object{NGC 985} \citep{Ebre16}, \object{NGC 3783} \citep{Mehd17}, \object{Mrk 817} \citep{Kara21}, and \object{NGC 3227} \citep{Mehd21a}, where the optical/UV continuum does not become dimmer during the obscuration events. This supports our modeling of the optical/UV dimming in \ngc as a change in the intrinsic continuum.

{The diminished brightness of the X-ray spectrum is a common characteristic of both \ngc and the obscured AGN, showing transformation into the spectrum of a typical Seyfert-2 galaxy, such as \object{NGC 1068} (e.g. \citealt{Graf21}). However, in the obscuration-event AGN their obscured and unobscured X-ray spectra are seen to converge towards hard X-rays, seen even in the case of strong obscuration in NGC~5548 \citep{Kaas14}. So the main difference between obscured and unobscured X-ray spectra is in the soft X-ray band where continuum absorption takes place, while the hard X-ray band is relatively unchanged. However, in \ngc the whole of the X-ray spectrum from soft to hard X-rays is significantly fainter than the historical observations. This suggests that the underlying broadband continuum in \ngc has become intrinsically dimmer. Our modeling shows that all continuum components of the \ngc SED have become fainter as a result of the changing-look phenomenon. This is a key difference to the obscuration-event AGN, where the underlying intrinsic SED does not show any major changes during the event (see e.g. \citealt{Meh15a,Mehd17}).

The behavior of AGN emission lines is another interesting distinction between \ngc and the obscuration-event AGN. In \ngc we find that the intrinsic luminosities of the \FeKa line, the \ion{O}{7} forbidden line, and the broad \ion{O}{8} Ly$\alpha$ line, that were previously seen in the \xmm EPIC-pn and RGS spectra \citep{Meh10}, have significantly dropped in 2017 as described in Section \ref{sect_abs}. This observed dimming of the X-ray emission lines is consistent with the changing-look behavior of the optical BLR emission lines in \ngc \citep{Shap19,Feng21,Okny21}, such as the clear order-of-magnitude drop in the flux of the H$\beta$ line \citep{Shap19}. This is in stark contrast to the behavior of the obscuration-event AGN, where the emission lines remain unchanged. In the transient obscuration-event AGN, where obscuration is likely only confined to our line of sight, the ionizing radiation that illuminates the BLR and the narrow-line region (NLR) is not attenuated, therefore the strength of the optical and X-ray emission lines is not changed. This can be seen in the case of transient obscuration event in NGC~3783, where the X-ray emission lines stand prominently above the absorbed X-ray continuum in the \xmm/RGS spectrum \citep{Mao19}. In the extreme case of NGC~5548, where long-lasting and heavy obscuration has been persistently present for about a decade, some impact on the AGN emission lines have been predicted if there is global obscuration in all directions \citep{Dehg19b,Dehg21}. However, the observed optical and X-ray emission lines in NGC~5548 have not noticeably changed \citep{Kaas14,Meh15a}, certainly not to the extent of \ngc, and also the predicted line changes in NGC~5548 are less than the major line changes seen in \ngc. Therefore, in the case of \ngc, it is most likely that the intrinsic dimming of the ionizing continuum results in lower ionization and less reprocessing by the line-emitting regions.
 
As shown in Table \ref{table_fit}, there is a significant drop in the normalization and the seed photon temperature $T_{\rm seed}$ of the Comptonized disk component ({\tt comt}) in 2017 compared to 2006. The electron temperature $T_{\rm e}$ of the corona remains unchanged between the two epochs. The change in the {\tt comt} component can be seen in Fig. \ref{fig_SED} (top right panel), where the luminosity of the thermal optical/UV emission from the disk is significantly lower during the 2017 observations. The dimming of the {\tt comt} component also results in a much weaker soft X-ray excess component. The intrinsic luminosity of the soft X-ray excess over 0.3--1.5 keV was ${6.7 \times 10^{42}}$~\ergs in 2006, while it is significantly lower by a factor of 9 (Obs. 1) and 8 (Obs. 2) in 2017. Such major changes in the strength of the soft X-ray excess have previously been shown to be a key characteristic of changing-look AGN, such as Mrk 1018 \citep{Noda18}. The soft X-ray excess, and its associated EUV emission, significantly contribute to the ionizing photons, therefore their vanishing or re-emergence is shown to have a major impact on the appearance of the BLR lines in changing-look AGN \citep{Noda18}. Interestingly, the \swift monitoring (Fig \ref{fig_lc}) shows that in 2020, when \ngc recovers to the high-flux state, as the UV flux peaks (top panel) the X-ray hardness ratio dips (bottom panel). Such X-ray spectral softening with UV brightening is a signature of the soft X-ray excess being produced by warm Comptonization, where the UV disk component and its emission tail at higher energies (i.e. the soft X-ray excess) become brighter together. This variability relation between the UV continuum and the soft X-ray excess has also been seen in other AGN, most notably \object{Mrk 509} \citep{Meh11}.

The intrinsic X-ray power-law luminosity of \ngc over 0.3--10 keV (${1.7 \times 10^{43}}$~\ergs in 2006) dropped by a factor of 10 (Obs. 1) and 4 (Obs. 2) in 2017. Also, the modeling results (Table \ref{table_fit}) suggest that the intrinsic power-law continuum is slightly harder in the 2017 changing-look state. Furthermore, there is a decline in the reflection luminosity between the two epochs as shown in Fig. \ref{fig_SED} (bottom right panel) and in Table \ref{table_lum}. This can be a manifestation of the primary X-ray continuum becoming dimmer in 2017 (Fig. \ref{fig_SED} and Table \ref{table_lum}), hence producing less reprocessed emission. This behavior of the \FeKa line responding to the fading of the continuum has been seen before in other changing-look AGN, such as Mrk~1018 \citep{LaMa17}.

According to our broadband continuum modeling we find that the intrinsic bolometric luminosity of \ngc in its normal high-flux state in 2006 was ${1.2 \times 10^{44}}$~\ergs, which, as a result of change in the accretion activity, dropped in the low-flux state to ${1.5 \times 10^{43}}$~\ergs (Obs. 1) and ${2.7 \times 10^{43}}$~\ergs (Obs. 2). Taking into account the black hole mass of 2.5~$\times 10^{7}$~$M_{\odot}$ \citep{Bent15}, the intrinsic bolometric luminosity of \ngc corresponds to about 4\% of the Eddington luminosity in the normal high-flux state (2006), and in the changing-look low-flux state (2017) to 0.5\% (Obs. 1) and 0.9\% (Obs. 2) of the Eddington luminosity. The observed timescale of the change in the production of seed photons from the accretion disk of \ngc is indicative of being driven by thermal fluctuations from the disk (see e.g. \citealt{Kell09}), rather than structural changes in the disk.

%%%%%%%%%%%%%%%%%%%%%%%%%%%%%%%%%%%%%%%%%%%%%%%%%%%%%%%%%%%%%%%%%%%%%%%%%%%%%%%%%%%%%%%%%%%%%%%%%%%%%%%
\subsection{Effect of dimming continuum on ionized gas}
\label{sect_outflows}

\swift monitoring of \ngc shows significant X-ray hardness variability (Fig. \ref{fig_lc}, bottom panel). This is seen on both long timescales (between the high-flux and low-flux epochs), and on short timescales (over a few weeks). The characteristics of the variability in \ngc are similar to those found in AGN with `transient obscuration events', such as NGC~5548 \citep{Mehd16}, where obscuring outflows passing in our line of sight cause X-ray hardness variability. However, in the case of \ngc we find that the variability is driven by the intrinsic continuum as shown by our modeling in this paper. As a consequence of the continuum variability, the ionization parameters of the warm-absorber components vary too, resulting in variation in their absorption, which would then contribute to the observed X-ray hardness variability (Fig. \ref{fig_lc}). This can be seen by the transmission model in Fig. \ref{fig_trans}, which shows that the continuum absorption by the warm-absorber is stronger in 2017 than in 2006, and on weeks timescales is stronger in Obs. 1 than in Obs. 2. These changes in the ionization of the warm-absorber results in additional X-ray absorption and spectral hardness variability. Therefore, without taking into account the de-ionization of the warm-absorber, one may erroneously fit a low-flux X-ray spectrum and its variability by introducing new additional obscuring gas, which we find not to be necessary in \ngc.

In \citet{Okny21}, where the reverberation of the \ngc optical emission lines is studied, based on the variable appearance of the \swift X-ray lightcurve the authors suggest that there is variable X-ray obscuration in \ngc. However, such a conclusion requires broadband spectral modeling and photoionization modeling, using X-ray spectra taken with \chandra or \xmm, as well as hard X-ray spectral coverage with \nustar, which were not carried out in \citet{Okny21}. Therefore, the effect of the changing ionizing SED on the already-present warm-absorber in \ngc was not considered in \citet{Okny21}. Nonetheless, our paper and \citet{Okny21} are both in agreement that `additional absorption' is needed to explain the data of \ngc; the difference is that in our paper we show that this `additional absorption' is a consequence of the dimming of the SED and hence the enhanced absorption by the de-ionized warm-absorber (Fig. \ref{fig_trans}), whereas in \citet{Okny21} this effect is not considered and instead the `additional absorption' is attributed to variable obscuration in our line of sight.

Adopting the scenario where the observed change in the X-ray emission lines is caused by a change in the ionizing SED, implies that the light travel time between the central ionizing source and the X-ray reprocessing regions must be less than the spacing between the two epochs (i.e. Oct. 2006 and Dec. 2017). Therefore, the X-ray emission line regions are located within a distance ${r < 3.4}$~pc from the central source. In the case of \ion{O}{7} and \ion{O}{8} emission lines, using the definition of ionization parameter $\xi$, the corresponding limit on density $n_{\rm H}$ can be estimated as ${n_{\rm H} = L_{\rm ion} / \xi\, r^2}$, where the ionizing luminosity $L_{\rm ion}$ is obtained from the broadband continuum modeling; the ionization parameter $\xi$ is given by the {\tt pion} model at which ionic concentrations of \ion{O}{7} ($\log \xi = 0.9$) and \ion{O}{8} ($\log \xi = 1.7$) peak; and $r$ is already constrained as described above. Thus, ${r < 3.4}$~pc implies that ${n_{\rm H} > 7 \times 10^{4}}$~\den for \ion{O}{7}, and ${n_{\rm H} > 1 \times 10^{4}}$~\den for the \ion{O}{8} emission line region.

Since we find the warm-absorber responds to changes in the ionizing SED, we can put limits on its density $n_{\rm H}$ and distance $r$ from the ionizing source, using the recombination timescale $t_{\rm rec}$ and the ionization parameter $\xi$. The {\tt pion} photoionization model provides $t_{\rm rec}$ for each ion according to the definition of \citet{Bott00}. The distance is calculated as ${r = \sqrt{L_{\rm ion} / \xi\, n_{\rm H}}}$, where the ionizing luminosity $L_{\rm ion}$ is known from the SED modeling; the ionization parameter $\xi$ is already given by the {\tt pion} model; and $n_{\rm H}$ is constrained by the density-dependent $t_{\rm rec}$. According to our modeling of \ngc, the warm-absorber components respond to the ionizing SED between the 2006 and 2017 epochs (i.e. de-ionization as a consequence of changing-look SED), as well as between Obs. 1 and 2 (separated by three weeks) as the intrinsic continuum varies. Since $t_{\rm rec}$ has to be shorter than the spacing between these observations, this can be used to put limits on $n_{\rm H}$ and hence $r$. For each component we calculated $t_{\rm rec}$ for the strongest Oxygen ion with the highest ionic concentration. From the warm-absorber de-ionization between the 2006 and 2017 epochs we find: 
${n_{\rm H} > 930}$~\den and ${r < 3.4}$~pc for Comp 1; 
${n_{\rm H} > 280}$~\den and ${r < 12}$~pc for Comp 2; 
${n_{\rm H} > 290}$~\den and ${r < 64}$~pc for Comp 3. 
Furthermore, the warm-absorber change on shorter timescales between Obs. 1 and 2 can be used to place tighter limits: 
${n_{\rm H} > 1 \times 10^{6}}$~\den and ${r < 0.1}$~pc for Comp 1; 
${n_{\rm H} > 9 \times 10^{4}}$~\den and ${r < 0.8}$~pc for Comp 2; 
${n_{\rm H} > 5 \times 10^{4}}$~\den and ${r < 6}$~pc for Comp 3. These constraints on the distance of the warm-absorber, and the X-ray emission lines (Section \ref {sect_accretion}), are comparable to those found for the ionized regions in NGC~5548 \citep{Arav15,Ebre16b,Mao18}.

A drop in the ionizing radiation that illuminates the absorbing gas in AGN could result in the appearance of new absorption lines in the UV and X-ray spectra as a consequence of the change in the photoionization equilibrium. However, it is almost always not possible to detect such new features in X-ray high-resolution spectra due to the low signal-to-noise ratio of the diminished X-ray continuum. In contrast, in the UV band, with HST, such features are often detectable (e.g. \citealt{Kris19b}). In the case of \ngc, no HST grating observations were taken during the epoch of the low-flux state. The last HST/COS observation of \ngc was taken in January 2011. The \swift lightcurve of \ngc (Fig. \ref{fig_lc}) shows that in 2012 the optical/UV and X-ray fluxes were already dropping. While there are no \swift observations available over 2007--2011, it is possible that at the time of the 2011 HST observation, \ngc was at the early stage of the changing-look event. Interestingly, the warm-absorber study of the 2011 HST/COS spectrum by \citet{Dunn18} finds the appearance of new absorption troughs that were not present in previous UV spectra, as well as changes in the previously-seen absorption troughs. 

Historically, UV spectra of \ngc have shown considerable variability in the characteristics of the associated absorbing gas. Several decades ago, observations with the {\it International Ultraviolet Explorer} (IUE) when \ngc was in a high flux state not only showed variability, but also strong, broad, blue-shifted absorption in \ion{C}{4} \citep{Voit87}. Observations with the {\it Hopkins Ultraviolet Telescope} (HUT) in 1995 also caught \ngc in a bright state, but the broad \ion{C}{4} absorption troughs had disappeared \citep{Kris96a}.
\cite{Math97} suggested that the disappearance of the broad \ion{C}{4} absorption and the simultaneous diminution of the previously strong soft X-ray absorption was due to the expansion of the outflowing gas as it decreased in density and its ionization parameter rose. The broad absorption reappeared in 2000 as \ngc entered a low flux state, showing up as the `Component 5' \citep{Krae02}. These same troughs are present in the 2011 HST/COS spectrum \citep{Dunn18} in addition to a new highly blue-shifted `Component 9'. Although \citet{Dunn18} interpret these spectral changes as evolution of the outflows through bulk motion across our line of sight, the similarity to previous episodes of variability associated with significant changes in the flux state suggests that these spectral changes are primarily a change in the ionization state of the absorbing gas in the outflow.
%%%%%%%%%%%%%%%%%%%%%%%%%%%%%%%%%%%%%%%%%%%%%%%%%%%%%%%%%%%%%%%%%%%%%%%%%%%%%%%%%%%%%%%%%%%%%%%%%%%%%%%
%%%%%%%%%%%%%%%%%%%%%%%%%%%%%%%%%%%%%%%%%%%%%%%%%%%%%%%%%%%%%%%%%%%%%%%%%%%%%%%%%%%%%%%%%%%%%%%%%%%%%%%
%%%%%%%%%%%%%%%%%%%%%%%%%%%%%%%%%%%%%%%%%%%%%%%%%%%%%%%%%%%%%%%%%%%%%%%%%%%%%%%%%%%%%%%%%%%%%%%%%%%%%%%
\section{Conclusions} 
\label{sect_concl}
In this paper we studied the SED and the intrinsic X-ray absorption in \ngc at two epochs, corresponding to its normal high-flux state (2006) and its changing-look low-flux state (2017). We modeled data taken with \chandra, \nustar, \xmm, and \swift. The comparison of the modeling results for the two epochs enables us to ascertain how the accretion-powered radiation and the ionized outflows differ between the two states of \ngc. Compared to its normal high-flux state in 2006, we find the intrinsic bolometric luminosity has dropped from ${1.2 \times 10^{44}}$~\ergs by a factor of 4 to 8 during the timespan of our 2017 observations. This dimming of the multi-wavelength continuum is explained as a decline in the optical/UV seed photons from the accretion disk, which consequently translates into lower Compton up-scattering to X-rays. As a consequence of the dimming of the primary continuum, the reprocessed X-ray emission lines (\FeKa, \ion{O}{7} forbidden, and \ion{O}{8} Ly$\alpha$) have also become fainter in 2017. This implies that the corresponding X-ray emission line regions are located at $r < 3.4$~pc from the central source. The transformation of the ionizing continuum from high-flux to low-flux state has a major impact on the ionization state of the ionized warm-absorber outflows. The change in the luminosity and the shape of the SED significantly lowers the ionization parameters of the three components of the warm-absorber in \ngc. This consequently results in enhanced X-ray absorption in our line of sight. Thus, variation in the ionization of the warm-absorber components, caused by intrinsic continuum variability, has a significant contribution to the observed X-ray spectral variability of \ngc, which is seen by the \swift monitoring on both short (weeks) and long (years) timescales. The response of the warm-absorber components to the ionizing SED implies that they are located at ${r < 0.1}$~pc for the highest-ionization component, ${r < 0.8}$~pc for the mid-ionization component, and ${r < 6}$~pc for the lowest-ionization component.
%\bigskip
%%%%%%%%%%%%%%%%%%%%%%%%%%%%%%%%%%%%%%%%%%%%%%%%%%%%%%%%%%%%%%%%%%%%%%%%%%%%%%%%%%%%%%%%%%%%%%%%%%%%%%%
%%%%%%%%%%%%%%%%%%%%%%%%%%%%%%%%%%%%%%%%%%%%%%%%%%%%%%%%%%%%%%%%%%%%%%%%%%%%%%%%%%%%%%%%%%%%%%%%%%%%%%%
%%%%%%%%%%%%%%%%%%%%%%%%%%%%%%%%%%%%%%%%%%%%%%%%%%%%%%%%%%%%%%%%%%%%%%%%%%%%%%%%%%%%%%%%%%%%%%%%%%%%%%%
\begin{acknowledgments}
This research has made use of data obtained from the Chandra Data Archive, and software provided by the Chandra X-ray Center (CXC). This research has made use of data obtained with the \nustar mission, a project led by the California Institute of Technology (Caltech), managed by the Jet Propulsion Laboratory (JPL) and funded by NASA. This work is based on observations obtained with XMM-Newton, an ESA science mission with instruments and contributions directly funded by ESA Member States and the USA (NASA). Support for this work was provided by the National Aeronautics and Space Administration through Chandra Award Number 18103X issued by the Chandra X-ray Center, which is operated by the Smithsonian Astrophysical Observatory for and on behalf of the National Aeronautics Space Administration under contract NAS8-03060. SRON is supported financially by NWO, the Netherlands Organization for Scientific Research. We thank the anonymous referee for providing constructive comments and suggestions that improved the paper.
\end{acknowledgments}
%\bigskip
%%%%%%%%%%%%%%%%%%%%%%%%%%%%%%%%%%%%%%%%%%%%%%%%%%%%%%%%%%%%%%%%%%%%%%%%%%%%%%%%%%%%%%%%%%%%%%%%%%%%%%%
%%%%%%%%%%%%%%%%%%%%%%%%%%%%%%%%%%%%%%%%%%%%%%%%%%%%%%%%%%%%%%%%%%%%%%%%%%%%%%%%%%%%%%%%%%%%%%%%%%%%%%%
%%%%%%%%%%%%%%%%%%%%%%%%%%%%%%%%%%%%%%%%%%%%%%%%%%%%%%%%%%%%%%%%%%%%%%%%%%%%%%%%%%%%%%%%%%%%%%%%%%%%%%%
%\newpage
\facilities{CXO (LETG), NuSTAR, Swift, XMM}
\software{SPEX \citep{Kaa96,Kaas20}, HEASoft \citep{HEASoft}}
%%%%%%%%%%%%%%%%%%%%%%%%%%%%%%%%%%%%%%%%%%%%%%%%%%%%%%%%%%%%%%%%%%%%%%%%%%%%%%%%%%%%%%%%%%%%%%%%%%%%%%%
%%%%%%%%%%%%%%%%%%%%%%%%%%%%%%%%%%%%%%%%%%%%%%%%%%%%%%%%%%%%%%%%%%%%%%%%%%%%%%%%%%%%%%%%%%%%%%%%%%%%%%%
%%%%%%%%%%%%%%%%%%%%%%%%%%%%%%%%%%%%%%%%%%%%%%%%%%%%%%%%%%%%%%%%%%%%%%%%%%%%%%%%%%%%%%%%%%%%%%%%%%%%%%%          
%\bibliography{sample631}{}
\bibliographystyle{aasjournal}
\bibliography{references}{}

\begin{thebibliography}{}
\expandafter\ifx\csname natexlab\endcsname\relax\def\natexlab#1{#1}\fi
\providecommand{\url}[1]{\href{#1}{#1}}
\providecommand{\dodoi}[1]{doi:~\href{http://doi.org/#1}{\nolinkurl{#1}}}
\providecommand{\doeprint}[1]{\href{http://ascl.net/#1}{\nolinkurl{http://ascl.net/#1}}}
\providecommand{\doarXiv}[1]{\href{https://arxiv.org/abs/#1}{\nolinkurl{https://arxiv.org/abs/#1}}}

\bibitem[{{Arav} {et~al.}(2015){Arav}, {Chamberlain}, {Kriss}, {Kaastra},
  {Cappi}, {Mehdipour}, {Petrucci}, {Steenbrugge}, {Behar}, {Bianchi},
  {Boissay}, {Branduardi-Raymont}, {Costantini}, {Ely}, {Ebrero}, {di Gesu},
  {Harrison}, {Kaspi}, {Malzac}, {De Marco}, {Matt}, {Nandra}, {Paltani},
  {Peterson}, {Pinto}, {Ponti}, {Pozo Nu{\~n}ez}, {De Rosa}, {Seta}, {Ursini},
  {de Vries}, {Walton}, \& {Whewell}}]{Arav15}
{Arav}, N., {Chamberlain}, C., {Kriss}, G.~A., {et~al.} 2015, \aap, 577, A37,
  \dodoi{10.1051/0004-6361/201425302}

\bibitem[{{Bentz} \& {Katz}(2015)}]{Bent15}
{Bentz}, M.~C., \& {Katz}, S. 2015, \pasp, 127, 67, \dodoi{10.1086/679601}

\bibitem[{{Bentz} {et~al.}(2013){Bentz}, {Denney}, {Grier}, {Barth},
  {Peterson}, {Vestergaard}, {Bennert}, {Canalizo}, {De Rosa}, {Filippenko},
  {Gates}, {Greene}, {Li}, {Malkan}, {Pogge}, {Stern}, {Treu}, \&
  {Woo}}]{Ben13}
{Bentz}, M.~C., {Denney}, K.~D., {Grier}, C.~J., {et~al.} 2013, \apj, 767, 149,
  \dodoi{10.1088/0004-637X/767/2/149}

\bibitem[{{Bottorff} {et~al.}(2000){Bottorff}, {Korista}, \&
  {Shlosman}}]{Bott00}
{Bottorff}, M.~C., {Korista}, K.~T., \& {Shlosman}, I. 2000, \apj, 537, 134,
  \dodoi{10.1086/309006}

\bibitem[{{Brinkman} {et~al.}(2000){Brinkman}, {Gunsing}, {Kaastra}, {van der
  Meer}, {Mewe}, {Paerels}, {Raassen}, {van Rooijen}, {Br{\"a}uninger},
  {Burkert}, {Burwitz}, {Hartner}, {Predehl}, {Ness}, {Schmitt}, {Drake},
  {Johnson}, {Juda}, {Kashyap}, {Murray}, {Pease}, {Ratzlaff}, \&
  {Wargelin}}]{Brink00}
{Brinkman}, A.~C., {Gunsing}, C.~J.~T., {Kaastra}, J.~S., {et~al.} 2000, \apjl,
  530, L111, \dodoi{10.1086/312504}

\bibitem[{{Burrows} {et~al.}(2005){Burrows}, {Hill}, {Nousek}, {Kennea},
  {Wells}, {Osborne}, {Abbey}, {Beardmore}, {Mukerjee}, {Short}, {Chincarini},
  {Campana}, {Citterio}, {Moretti}, {Pagani}, {Tagliaferri}, {Giommi},
  {Capalbi}, {Tamburelli}, {Angelini}, {Cusumano}, {Br{\"a}uninger}, {Burkert},
  \& {Hartner}}]{Burr05}
{Burrows}, D.~N., {Hill}, J.~E., {Nousek}, J.~A., {et~al.} 2005, \ssr, 120,
  165, \dodoi{10.1007/s11214-005-5097-2}

\bibitem[{{Cardelli} {et~al.}(1989){Cardelli}, {Clayton}, \& {Mathis}}]{Car89}
{Cardelli}, J.~A., {Clayton}, G.~C., \& {Mathis}, J.~S. 1989, \apj, 345, 245,
  \dodoi{10.1086/167900}

\bibitem[{{Cash}(1979)}]{Cash79}
{Cash}, W. 1979, \apj, 228, 939, \dodoi{10.1086/156922}

\bibitem[{{Cohen} {et~al.}(1986){Cohen}, {Rudy}, {Puetter}, {Ake}, \&
  {Foltz}}]{Cohe86}
{Cohen}, R.~D., {Rudy}, R.~J., {Puetter}, R.~C., {Ake}, T.~B., \& {Foltz},
  C.~B. 1986, \apj, 311, 135, \dodoi{10.1086/164758}

\bibitem[{{Costantini} {et~al.}(2000){Costantini}, {Nicastro}, {Fruscione},
  {Mathur}, {Comastri}, , {et~al.}}]{Cos00}
{Costantini}, E., {Nicastro}, F., {Fruscione}, A., {et~al.} 2000, \apj, 544,
  283, \dodoi{10.1086/317200}

\bibitem[{{Crummy} {et~al.}(2006){Crummy}, {Fabian}, {Gallo}, \&
  {Ross}}]{Cru06}
{Crummy}, J., {Fabian}, A.~C., {Gallo}, L., \& {Ross}, R.~R. 2006, \mnras, 365,
  1067, \dodoi{10.1111/j.1365-2966.2005.09844.x}

\bibitem[{{de Plaa} {et~al.}(2004){de Plaa}, {Kaastra}, {Tamura},
  {Pointecouteau}, {Mendez}, \& {Peterson}}]{dePl04}
{de Plaa}, J., {Kaastra}, J.~S., {Tamura}, T., {et~al.} 2004, \aap, 423, 49,
  \dodoi{10.1051/0004-6361:20047170}

\bibitem[{{Dehghanian} {et~al.}(2019){Dehghanian}, {Ferland}, {Peterson},
  {Kriss}, {Korista}, {Chatzikos}, {Guzm{\'a}n}, {Arav}, {De Rosa}, {Goad},
  {Mehdipour}, \& {van Hoof}}]{Dehg19b}
{Dehghanian}, M., {Ferland}, G.~J., {Peterson}, B.~M., {et~al.} 2019, \apjl,
  882, L30, \dodoi{10.3847/2041-8213/ab3d41}

\bibitem[{{Dehghanian} {et~al.}(2021){Dehghanian}, {Ferland}, {Peterson},
  {Kriss}, {Korista}, {Goad}, {Chatzikos}, {Bentz}, {Guzm{\'a}n}, {Mehdipour},
  \& {De Rosa}}]{Dehg21}
---. 2021, \apj, 906, 14, \dodoi{10.3847/1538-4357/abcb91}

\bibitem[{{Denney} {et~al.}(2014){Denney}, {De Rosa}, {Croxall}, {Gupta},
  {Bentz}, {Fausnaugh}, {Grier}, {Martini}, {Mathur}, {Peterson}, {Pogge}, \&
  {Shappee}}]{Denn14}
{Denney}, K.~D., {De Rosa}, G., {Croxall}, K., {et~al.} 2014, \apj, 796, 134,
  \dodoi{10.1088/0004-637X/796/2/134}

\bibitem[{{Done} {et~al.}(2012){Done}, {Davis}, {Jin}, {Blaes}, \&
  {Ward}}]{Done12}
{Done}, C., {Davis}, S.~W., {Jin}, C., {Blaes}, O., \& {Ward}, M. 2012, \mnras,
  420, 1848, \dodoi{10.1111/j.1365-2966.2011.19779.x}

\bibitem[{{Dunn} {et~al.}(2018){Dunn}, {Parvaresh}, {Kraemer}, \&
  {Crenshaw}}]{Dunn18}
{Dunn}, J.~P., {Parvaresh}, R., {Kraemer}, S.~B., \& {Crenshaw}, D.~M. 2018,
  \apj, 854, 166, \dodoi{10.3847/1538-4357/aaa95d}

\bibitem[{{Ebrero} {et~al.}(2016{\natexlab{a}}){Ebrero}, {Kriss}, {Kaastra}, \&
  {Ely}}]{Ebre16}
{Ebrero}, J., {Kriss}, G.~A., {Kaastra}, J.~S., \& {Ely}, J.~C.
  2016{\natexlab{a}}, \aap, 586, A72, \dodoi{10.1051/0004-6361/201527495}

\bibitem[{{Ebrero} {et~al.}(2016{\natexlab{b}}){Ebrero}, {Kaastra}, {Kriss},
  {Di Gesu}, {Costantini}, {Mehdipour}, {Bianchi}, {Cappi}, {Boissay},
  {Branduardi-Raymont}, {Petrucci}, {Ponti}, {Pozo N{\'u}{\~n}ez}, {Seta},
  {Steenbrugge}, \& {Whewell}}]{Ebre16b}
{Ebrero}, J., {Kaastra}, J.~S., {Kriss}, G.~A., {et~al.} 2016{\natexlab{b}},
  \aap, 587, A129, \dodoi{10.1051/0004-6361/201527808}

\bibitem[{{Elitzur} {et~al.}(2014){Elitzur}, {Ho}, \& {Trump}}]{Elit14}
{Elitzur}, M., {Ho}, L.~C., \& {Trump}, J.~R. 2014, \mnras, 438, 3340,
  \dodoi{10.1093/mnras/stt2445}

\bibitem[{{Evans} {et~al.}(2009){Evans}, {Beardmore}, {Page}, {Osborne},
  {O'Brien}, , {et~al.}}]{Eva09}
{Evans}, P.~A., {Beardmore}, A.~P., {Page}, K.~L., {et~al.} 2009, \mnras, 397,
  1177, \dodoi{10.1111/j.1365-2966.2009.14913.x}

\bibitem[{{Evans} {et~al.}(2007){Evans}, {Beardmore}, {Page}, {Tyler},
  {Osborne}, {Goad}, {O'Brien}, {Vetere}, {Racusin}, {Morris}, {Burrows},
  {Capalbi}, {Perri}, {Gehrels}, \& {Romano}}]{Evan07}
---. 2007, \aap, 469, 379, \dodoi{10.1051/0004-6361:20077530}

\bibitem[{{Feng} {et~al.}(2021){Feng}, {Hu}, {Li}, {Liu}, {Bai}, {Xing},
  {Wang}, {Yang}, {Xiao}, \& {Lu}}]{Feng21}
{Feng}, H.-C., {Hu}, C., {Li}, S.-S., {et~al.} 2021, \apj, 909, 18,
  \dodoi{10.3847/1538-4357/abd851}

\bibitem[{{Ferrarese} \& {Merritt}(2000)}]{Ferr00}
{Ferrarese}, L., \& {Merritt}, D. 2000, \apjl, 539, L9, \dodoi{10.1086/312838}

\bibitem[{{Fruscione} {et~al.}(2006){Fruscione}, {McDowell}, {Allen},
  {Brickhouse}, {Burke}, {Davis}, {Durham}, {Elvis}, {Galle}, {Harris},
  {Huenemoerder}, {Houck}, {Ishibashi}, {Karovska}, {Nicastro}, {Noble},
  {Nowak}, {Primini}, {Siemiginowska}, {Smith}, \& {Wise}}]{Frus06}
{Fruscione}, A., {McDowell}, J.~C., {Allen}, G.~E., {et~al.} 2006, in
  \procspie, Vol. 6270, Society of Photo-Optical Instrumentation Engineers
  (SPIE) Conference Series, 62701V, \dodoi{10.1117/12.671760}

\bibitem[{{Gehrels} {et~al.}(2004){Gehrels}, {Chincarini}, {Giommi}, {Mason},
  {Nousek}, {Wells}, {White}, {Barthelmy}, {Burrows}, {Cominsky}, {Hurley},
  {Marshall}, {M{\'e}sz{\'a}ros}, {Roming}, {Angelini}, {Barbier}, {Belloni},
  {Campana}, {Caraveo}, {Chester}, {Citterio}, {Cline}, {Cropper}, {Cummings},
  {Dean}, {Feigelson}, {Fenimore}, {Frail}, {Fruchter}, {Garmire}, {Gendreau},
  {Ghisellini}, {Greiner}, {Hill}, {Hunsberger}, {Krimm}, {Kulkarni}, {Kumar},
  {Lebrun}, {Lloyd-Ronning}, {Markwardt}, {Mattson}, {Mushotzky}, {Norris},
  {Osborne}, {Paczynski}, {Palmer}, {Park}, {Parsons}, {Paul}, {Rees},
  {Reynolds}, {Rhoads}, {Sasseen}, {Schaefer}, {Short}, {Smale}, {Smith},
  {Stella}, {Tagliaferri}, {Takahashi}, {Tashiro}, {Townsley}, {Tueller},
  {Turner}, {Vietri}, {Voges}, {Ward}, {Willingale}, {Zerbi}, \&
  {Zhang}}]{Gehr04}
{Gehrels}, N., {Chincarini}, G., {Giommi}, P., {et~al.} 2004, \apj, 611, 1005,
  \dodoi{10.1086/422091}

\bibitem[{{Grafton-Waters} {et~al.}(2021){Grafton-Waters},
  {Branduardi-Raymont}, {Mehdipour}, {Page}, {Bianchi}, {Behar}, \&
  {Symeonidis}}]{Graf21}
{Grafton-Waters}, S., {Branduardi-Raymont}, G., {Mehdipour}, M., {et~al.} 2021,
  \aap, 649, A162, \dodoi{10.1051/0004-6361/202039022}

\bibitem[{{Harrison} {et~al.}(2013){Harrison}, {Craig}, {Christensen},
  {Hailey}, {Zhang}, {Boggs}, {Stern}, {Cook}, {Forster}, {Giommi},
  {Grefenstette}, {Kim}, {Kitaguchi}, {Koglin}, {Madsen}, {Mao}, {Miyasaka},
  {Mori}, {Perri}, {Pivovaroff}, {Puccetti}, {Rana}, {Westergaard}, {Willis},
  {Zoglauer}, {An}, {Bachetti}, {Barri{\`e}re}, {Bellm}, {Bhalerao},
  {Brejnholt}, {Fuerst}, {Liebe}, {Markwardt}, {Nynka}, {Vogel}, {Walton},
  {Wik}, {Alexander}, {Cominsky}, {Hornschemeier}, {Hornstrup}, {Kaspi},
  {Madejski}, {Matt}, {Molendi}, {Smith}, {Tomsick}, {Ajello}, {Ballantyne},
  {Balokovi{\'c}}, {Barret}, {Bauer}, {Blandford}, {Brandt}, {Brenneman},
  {Chiang}, {Chakrabarty}, {Chenevez}, {Comastri}, {Dufour}, {Elvis}, {Fabian},
  {Farrah}, {Fryer}, {Gotthelf}, {Grindlay}, {Helfand}, {Krivonos}, {Meier},
  {Miller}, {Natalucci}, {Ogle}, {Ofek}, {Ptak}, {Reynolds}, {Rigby},
  {Tagliaferri}, {Thorsett}, {Treister}, \& {Urry}}]{Harr13}
{Harrison}, F.~A., {Craig}, W.~W., {Christensen}, F.~E., {et~al.} 2013, \apj,
  770, 103, \dodoi{10.1088/0004-637X/770/2/103}

\bibitem[{{Holczer} \& {Behar}(2012)}]{Holc12}
{Holczer}, T., \& {Behar}, E. 2012, \apj, 747, 71,
  \dodoi{10.1088/0004-637X/747/1/71}

\bibitem[{{Huerta} {et~al.}(2014){Huerta}, {Krongold}, {Nicastro}, {Mathur},
  {Longinotti}, \& {Jimenez-Bailon}}]{Huer14}
{Huerta}, E.~M., {Krongold}, Y., {Nicastro}, F., {et~al.} 2014, \apj, 793, 61,
  \dodoi{10.1088/0004-637X/793/1/61}

\bibitem[{{Ili{\'c}} {et~al.}(2020){Ili{\'c}}, {Oknyansky}, {Popovi{\'c}},
  {Tsygankov}, {Belinski}, {Tatarnikov}, {Dodin}, {Shatsky}, {Ikonnikova},
  {Raki{\'c}}, {Kova{\v{c}}evi{\'c}}, {Mar{\v{c}}eta-Mandi{\'c}}, {Burlak},
  {Mishin}, {Metlova}, {Potanin}, \& {Zheltoukhov}}]{Ilic20}
{Ili{\'c}}, D., {Oknyansky}, V., {Popovi{\'c}}, L.~{\v{C}}., {et~al.} 2020,
  \aap, 638, A13, \dodoi{10.1051/0004-6361/202037532}

\bibitem[{{Kaastra}(2017)}]{Kaas17}
{Kaastra}, J.~S. 2017, \aap, 605, A51, \dodoi{10.1051/0004-6361/201629319}

\bibitem[{{Kaastra} \& {Bleeker}(2016)}]{Kaas16}
{Kaastra}, J.~S., \& {Bleeker}, J.~A.~M. 2016, \aap, 587, A151,
  \dodoi{10.1051/0004-6361/201527395}

\bibitem[{{Kaastra} {et~al.}(1996){Kaastra}, {Mewe}, \&
  {Nieuwenhuijzen}}]{Kaa96}
{Kaastra}, J.~S., {Mewe}, R., \& {Nieuwenhuijzen}, H. 1996, in UV and X-ray
  Spectroscopy of Astrophysical and Laboratory Plasmas, ed. K.~{Yamashita} \&
  T.~{Watanabe}, 411--414

\bibitem[{Kaastra {et~al.}(2020)Kaastra, Raassen, de~Plaa, \& Gu}]{Kaas20}
Kaastra, J.~S., Raassen, A. J.~J., de~Plaa, J., \& Gu, L. 2020, SPEX X-ray
  spectral fitting package, 3.06.01,  Zenodo, \dodoi{10.5281/zenodo.4384188}

\bibitem[{Kaastra {et~al.}(2014)Kaastra, Kriss, Cappi, Mehdipour, Petrucci,
  Steenbrugge, Arav, Behar, Bianchi, Boissay, Branduardi-Raymont, Chamberlain,
  Costantini, Ely, Ebrero, Di~Gesu, Harrison, Kaspi, Malzac, De~Marco, Matt,
  Nandra, Paltani, Person, Peterson, Pinto, Ponti, Nuñez, De~Rosa, Seta,
  Ursini, de~Vries, Walton, \& Whewell}]{Kaas14}
Kaastra, J.~S., Kriss, G.~A., Cappi, M., {et~al.} 2014, Science, 345, 64,
  \dodoi{10.1126/science.1253787}

\bibitem[{{Kara} {et~al.}(2021){Kara}, {Mehdipour}, {Kriss}, {Cackett}, {Arav},
  {Barth}, {Byun}, {Brotherton}, {De Rosa}, {Gelbord}, {Hernandez Santisteban},
  {Hu}, {Kaastra}, {Landt}, {Li}, {Miller}, {Montano}, {Partington},
  {Aceituno}, {Bai}, {Bao}, {Bentz}, {Brink}, {Chelouche}, {Chen}, {Dalla
  Bonta}, {Dehghanian}, {Du}, {Edelson}, {Ferland}, {Ferrarese}, {Fian},
  {Filippenko}, {Fischer}, {Goad}, {Gonzalez Buitrago}, {Gorjian}, {Grier},
  {Guo}, {Hall}, {Homayouni}, {Horne}, {Ilic}, {Jiang}, {Joner}, {Kaspi},
  {Kochanek}, {Korista}, {Kynoch}, {Li}, {Liu}, {Mc Hardy}, {McLane},
  {Mitchell}, {Netzer}, {Olson}, {Pogge}, {Popovic}, {Proga},
  {Storchi-Bergmann}, {Strasburger}, {Treu}, {Vestergaard}, {Wang}, {Ward},
  {Waters}, {Williams}, {Yang}, {Yao}, {Zastrocky}, {Zhai}, \& {Zu}}]{Kara21}
{Kara}, E., {Mehdipour}, M., {Kriss}, G.~A., {et~al.} 2021, \apj, submitted,
  arXiv:2105.05840.
\newblock \doarXiv{2105.05840}

\bibitem[{{Keel}(1996)}]{Keel96}
{Keel}, W.~C. 1996, \aj, 111, 696, \dodoi{10.1086/117816}

\bibitem[{{Kelly} {et~al.}(2009){Kelly}, {Bechtold}, \&
  {Siemiginowska}}]{Kell09}
{Kelly}, B.~C., {Bechtold}, J., \& {Siemiginowska}, A. 2009, \apj, 698, 895,
  \dodoi{10.1088/0004-637X/698/1/895}

\bibitem[{{Kinney} {et~al.}(1996){Kinney}, {Calzetti}, {Bohlin}, {McQuade},
  {Storchi-Bergmann}, , {et~al.}}]{Kin96}
{Kinney}, A.~L., {Calzetti}, D., {Bohlin}, R.~C., {et~al.} 1996, \apj, 467, 38,
  \dodoi{10.1086/177583}

\bibitem[{{Komossa}(2015)}]{Komo15}
{Komossa}, S. 2015, Journal of High Energy Astrophysics, 7, 148,
  \dodoi{10.1016/j.jheap.2015.04.006}

\bibitem[{{Kraemer} {et~al.}(2002){Kraemer}, {Crenshaw}, {George}, {Netzer},
  {Turner}, \& {Gabel}}]{Krae02}
{Kraemer}, S.~B., {Crenshaw}, D.~M., {George}, I.~M., {et~al.} 2002, \apj, 577,
  98, \dodoi{10.1086/342173}

\bibitem[{{Kriss} {et~al.}(1996{\natexlab{a}}){Kriss}, {Espey}, {Krolik},
  {Tsvetanov}, {Zheng}, \& {Davidsen}}]{Kris96a}
{Kriss}, G.~A., {Espey}, B.~R., {Krolik}, J.~H., {et~al.} 1996{\natexlab{a}},
  \apj, 467, 622, \dodoi{10.1086/177637}

\bibitem[{{Kriss} {et~al.}(1996{\natexlab{b}}){Kriss}, {Krolik}, {Otani},
  {Espey}, {Turner}, {Kii}, {Tsvetanov}, {Takahashi}, {Davidsen}, {Tashiro},
  {Zheng}, {Murakami}, {Petre}, \& {Mihara}}]{Kris96b}
{Kriss}, G.~A., {Krolik}, J.~H., {Otani}, C., {et~al.} 1996{\natexlab{b}},
  \apj, 467, 629, \dodoi{10.1086/177638}

\bibitem[{{Kriss} {et~al.}(2019{\natexlab{a}}){Kriss}, {Mehdipour}, {Kaastra},
  {Rau}, {Bodensteiner}, {Plesha}, {Arav}, {Behar}, {Bianchi},
  {Branduardi-Raymont}, {Cappi}, {Costantini}, {De Marco}, {Di Gesu}, {Ebrero},
  {Kaspi}, {Mao}, {Middei}, {Miller}, {Paltani}, {Peretz}, {Peterson},
  {Petrucci}, {Ponti}, {Ursini}, {Walton}, \& {Xu}}]{Kris19}
{Kriss}, G.~A., {Mehdipour}, M., {Kaastra}, J.~S., {et~al.} 2019{\natexlab{a}},
  \aap, 621, A12, \dodoi{10.1051/0004-6361/201834326}

\bibitem[{{Kriss} {et~al.}(2019{\natexlab{b}}){Kriss}, {De Rosa}, {Ely},
  {Peterson}, {Kaastra}, {Mehdipour}, {Ferland}, {Dehghanian}, {Mathur},
  {Edelson}, {Korista}, {Arav}, {Barth}, {Bentz}, {Brandt}, {Crenshaw}, {Dalla
  Bont{\`a}}, {Denney}, {Done}, {Eracleous}, {Fausnaugh}, {Gardner}, {Goad},
  {Grier}, {Horne}, {Kochanek}, {McHardy}, {Netzer}, {Pancoast}, {Pei},
  {Pogge}, {Proga}, {Silva}, {Tejos}, {Vestergaard}, {Adams}, {Anderson},
  {Ar{\'e}valo}, {Beatty}, {Behar}, {Bennert}, {Bianchi}, {Bigley}, {Bisogni},
  {Boissay-Malaquin}, {Borman}, {Bottorff}, {Breeveld}, {Brotherton}, {Brown},
  {Brown}, {Cackett}, {Canalizo}, {Cappi}, {Carini}, {Clubb}, {Comerford},
  {Coker}, {Corsini}, {Costantini}, {Croft}, {Croxall}, {Deason}, {De
  Lorenzo-C{\'a}ceres}, {De Marco}, {Dietrich}, {Di Gesu}, {Ebrero}, {Evans},
  {Filippenko}, {Flatland}, {Gates}, {Gehrels}, {Geier}, {Gelbord}, {Gonzalez},
  {Gorjian}, {Grupe}, {Gupta}, {Hall}, {Henderson}, {Hicks}, {Holmbeck},
  {Holoien}, {Hutchison}, {Im}, {Jensen}, {Johnson}, {Joner}, {Kaspi}, {Kelly},
  {Kelly}, {Kennea}, {Kim}, {Kim}, {Kim}, {King}, {Klimanov}, {Krongold},
  {Lau}, {Lee}, {Leonard}, {Li}, {Lira}, {Lochhaas}, {Ma}, {MacInnis},
  {Malkan}, {Manne-Nicholas}, {Matt}, {Mauerhan}, {McGurk}, {Montuori},
  {Morelli}, {Mosquera}, {Mudd}, {M{\"u}ller-S{\'a}nchez}, {Nazarov}, {Norris},
  {Nousek}, {Nguyen}, {Ochner}, {Okhmat}, {Paltani}, {Parks}, {Pinto},
  {Pizzella}, {Poleski}, {Ponti}, {Pott}, {Rafter}, {Rix}, {Runnoe}, {Saylor},
  {Schimoia}, {Schn{\"u}lle}, {Scott}, {Sergeev}, {Shappee}, {Shivvers},
  {Siegel}, {Simonian}, {Siviero}, {Skielboe}, {Somers}, {Spencer}, {Starkey},
  {Stevens}, {Sung}, {Tayar}, {Teems}, {Treu}, {Turner}, {Uttley}, {. Van
  Saders}, {Vican}, {Villforth}, {Villanueva}, {Walton}, {Waters}, {Weiss},
  {Woo}, {Yan}, {Yuk}, {Zheng}, {Zhu}, \& {Zu}}]{Kris19b}
{Kriss}, G.~A., {De Rosa}, G., {Ely}, J., {et~al.} 2019{\natexlab{b}}, \apj,
  881, 153, \dodoi{10.3847/1538-4357/ab3049}

\bibitem[{{Krolik} {et~al.}(1981){Krolik}, {McKee}, \& {Tarter}}]{Kro81}
{Krolik}, J.~H., {McKee}, C.~F., \& {Tarter}, C.~B. 1981, \apj, 249, 422,
  \dodoi{10.1086/159303}

\bibitem[{{Kubota} \& {Done}(2018)}]{Kubo18}
{Kubota}, A., \& {Done}, C. 2018, \mnras, 480, 1247,
  \dodoi{10.1093/mnras/sty1890}

\bibitem[{{Laha} {et~al.}(2021){Laha}, {Reynolds}, {Reeves}, {Kriss},
  {Guainazzi}, {Smith}, {Veilleux}, \& {Proga}}]{Laha21}
{Laha}, S., {Reynolds}, C.~S., {Reeves}, J., {et~al.} 2021, Nature Astronomy,
  5, 13, \dodoi{10.1038/s41550-020-01255-2}

\bibitem[{{LaMassa} {et~al.}(2017){LaMassa}, {Yaqoob}, \& {Kilgard}}]{LaMa17}
{LaMassa}, S.~M., {Yaqoob}, T., \& {Kilgard}, R. 2017, \apj, 840, 11,
  \dodoi{10.3847/1538-4357/aa68df}

\bibitem[{{LaMassa} {et~al.}(2015){LaMassa}, {Cales}, {Moran}, {Myers},
  {Richards}, {Eracleous}, {Heckman}, {Gallo}, \& {Urry}}]{LaMa15}
{LaMassa}, S.~M., {Cales}, S., {Moran}, E.~C., {et~al.} 2015, \apj, 800, 144,
  \dodoi{10.1088/0004-637X/800/2/144}

\bibitem[{{Lodders} {et~al.}(2009){Lodders}, {Palme}, \& {Gail}}]{Lod09}
{Lodders}, K., {Palme}, H., \& {Gail}, H.-P. 2009, Landolt B{\"o}rnstein, 44,
  \dodoi{10.1007/978-3-540-88055-4_34}

\bibitem[{{Longinotti} {et~al.}(2013){Longinotti}, {Krongold}, {Kriss}, {Ely},
  {Gallo}, {Grupe}, {Komossa}, {Mathur}, \& {Pradhan}}]{Long13}
{Longinotti}, A.~L., {Krongold}, Y., {Kriss}, G.~A., {et~al.} 2013, \apj, 766,
  104, \dodoi{10.1088/0004-637X/766/2/104}

\bibitem[{{MacLeod} {et~al.}(2016){MacLeod}, {Ross}, {Lawrence}, {Goad},
  {Horne}, {Burgett}, {Chambers}, {Flewelling}, {Hodapp}, {Kaiser}, {Magnier},
  {Wainscoat}, \& {Waters}}]{MacL16}
{MacLeod}, C.~L., {Ross}, N.~P., {Lawrence}, A., {et~al.} 2016, \mnras, 457,
  389, \dodoi{10.1093/mnras/stv2997}

\bibitem[{{Magdziarz} \& {Zdziarski}(1995)}]{Magd95}
{Magdziarz}, P., \& {Zdziarski}, A.~A. 1995, \mnras, 273, 837

\bibitem[{{Mao} {et~al.}(2018){Mao}, {Kaastra}, {Mehdipour}, {Gu},
  {Costantini}, {Kriss}, {Bianchi}, {Branduardi-Raymont}, {Behar}, {Di Gesu},
  {Ponti}, {Petrucci}, \& {Ebrero}}]{Mao18}
{Mao}, J., {Kaastra}, J.~S., {Mehdipour}, M., {et~al.} 2018, \aap, 612, A18,
  \dodoi{10.1051/0004-6361/201732162}

\bibitem[{{Mao} {et~al.}(2019){Mao}, {Mehdipour}, {Kaastra}, {Costantini},
  {Pinto}, {Branduardi-Raymont}, {Behar}, {Peretz}, {Bianchi}, {Kriss},
  {Ponti}, {De Marco}, {Petrucci}, {Di Gesu}, {Middei}, {Ebrero}, \&
  {Arav}}]{Mao19}
{Mao}, J., {Mehdipour}, M., {Kaastra}, J.~S., {et~al.} 2019, \aap, 621, A99,
  \dodoi{10.1051/0004-6361/201833191}

\bibitem[{{Markowitz} {et~al.}(2008){Markowitz}, {Reeves}, {Miniutti},
  {Serlemitsos}, {Kunieda}, , {et~al.}}]{Mar08}
{Markowitz}, A., {Reeves}, J.~N., {Miniutti}, G., {et~al.} 2008, \pasj, 60,
  277.
\newblock \doarXiv{0710.0382}

\bibitem[{{Mathur} {et~al.}(1997){Mathur}, {Wilkes}, \& {Aldcroft}}]{Math97}
{Mathur}, S., {Wilkes}, B.~J., \& {Aldcroft}, T. 1997, \apj, 478, 182,
  \dodoi{10.1086/303761}

\bibitem[{{Matt} {et~al.}(2003){Matt}, {Guainazzi}, \& {Maiolino}}]{Matt03}
{Matt}, G., {Guainazzi}, M., \& {Maiolino}, R. 2003, \mnras, 342, 422,
  \dodoi{10.1046/j.1365-8711.2003.06539.x}

\bibitem[{{Mehdipour} {et~al.}(2011){Mehdipour}, {Branduardi-Raymont},
  {Kaastra}, {Petrucci}, {Kriss}, , {et~al.}}]{Meh11}
{Mehdipour}, M., {Branduardi-Raymont}, G., {Kaastra}, J.~S., {et~al.} 2011,
  \aap, 534, A39, \dodoi{10.1051/0004-6361/201116875}

\bibitem[{{Mehdipour} {et~al.}(2010){Mehdipour}, {Branduardi-Raymont}, \&
  {Page}}]{Meh10}
{Mehdipour}, M., {Branduardi-Raymont}, G., \& {Page}, M.~J. 2010, \aap, 514,
  A100, \dodoi{10.1051/0004-6361/200913049}

\bibitem[{{Mehdipour} {et~al.}(2016{\natexlab{a}}){Mehdipour}, {Kaastra}, \&
  {Kallman}}]{Meh16b}
{Mehdipour}, M., {Kaastra}, J.~S., \& {Kallman}, T. 2016{\natexlab{a}}, \aap,
  596, A65, \dodoi{10.1051/0004-6361/201628721}

\bibitem[{{Mehdipour} {et~al.}(2015){Mehdipour}, {Kaastra}, {Kriss}, {Cappi},
  {Petrucci}, {Steenbrugge}, {Arav}, {Behar}, {Bianchi}, {Boissay},
  {Branduardi-Raymont}, {Costantini}, {Ebrero}, {Di Gesu}, {Harrison}, {Kaspi},
  {De Marco}, {Matt}, {Paltani}, {Peterson}, {Ponti}, {Pozo Nu{\~n}ez}, {De
  Rosa}, {Ursini}, {de Vries}, {Walton}, \& {Whewell}}]{Meh15a}
{Mehdipour}, M., {Kaastra}, J.~S., {Kriss}, G.~A., {et~al.} 2015, \aap, 575,
  A22, \dodoi{10.1051/0004-6361/201425373}

\bibitem[{{Mehdipour} {et~al.}(2016{\natexlab{b}}){Mehdipour}, {Kaastra},
  {Kriss}, {Cappi}, {Petrucci}, {De Marco}, {Ponti}, {Steenbrugge}, {Behar},
  {Bianchi}, {Branduardi-Raymont}, {Costantini}, {Ebrero}, {Di Gesu}, {Matt},
  {Paltani}, {Peterson}, {Ursini}, \& {Whewell}}]{Mehd16}
---. 2016{\natexlab{b}}, \aap, 588, A139, \dodoi{10.1051/0004-6361/201527729}

\bibitem[{{Mehdipour} {et~al.}(2017){Mehdipour}, {Kaastra}, {Kriss}, {Arav},
  {Behar}, {Bianchi}, {Branduardi-Raymont}, {Cappi}, {Costantini}, {Ebrero},
  {Di Gesu}, {Kaspi}, {Mao}, {De Marco}, {Matt}, {Paltani}, {Peretz},
  {Peterson}, {Petrucci}, {Pinto}, {Ponti}, {Ursini}, {de Vries}, \&
  {Walton}}]{Mehd17}
---. 2017, \aap, 607, A28, \dodoi{10.1051/0004-6361/201731175}

\bibitem[{{Mehdipour} {et~al.}(2021){Mehdipour}, {Kriss}, {Kaastra}, {Wang},
  {Mao}, {Costantini}, {Arav}, {Behar}, {Bianchi}, {Branduardi-Raymont},
  {Brotherton}, {Cappi}, {De Marco}, {Di Gesu}, {Ebrero}, {Grafton-Waters},
  {Kaspi}, {Matt}, {Paltani}, {Petrucci}, {Pinto}, {Ponti}, {Ursini}, \&
  {Walton}}]{Mehd21a}
{Mehdipour}, M., {Kriss}, G.~A., {Kaastra}, J.~S., {et~al.} 2021, arXiv
  e-prints, arXiv:2106.14957.
\newblock \doarXiv{2106.14957}

\bibitem[{{Merloni} {et~al.}(2015){Merloni}, {Dwelly}, {Salvato},
  {Georgakakis}, {Greiner}, {Krumpe}, {Nandra}, {Ponti}, \& {Rau}}]{Merl15}
{Merloni}, A., {Dwelly}, T., {Salvato}, M., {et~al.} 2015, \mnras, 452, 69,
  \dodoi{10.1093/mnras/stv1095}

\bibitem[{{Miller} {et~al.}(2015){Miller}, {Kaastra}, {Miller}, {Reynolds},
  {Brown}, {Cenko}, {Drake}, {Gezari}, {Guillochon}, {Gultekin}, {Irwin},
  {Levan}, {Maitra}, {Maksym}, {Mushotzky}, {O'Brien}, {Paerels}, {de Plaa},
  {Ramirez-Ruiz}, {Strohmayer}, \& {Tanvir}}]{Mill15}
{Miller}, J.~M., {Kaastra}, J.~S., {Miller}, M.~C., {et~al.} 2015, \nat, 526,
  542, \dodoi{10.1038/nature15708}

\bibitem[{{Nasa High Energy Astrophysics Science Archive Research Center
  (Heasarc)}(2014)}]{HEASoft}
{Nasa High Energy Astrophysics Science Archive Research Center (Heasarc)}.
  2014, {HEAsoft: Unified Release of FTOOLS and XANADU}.
\newblock \doeprint{1408.004}

\bibitem[{{Netzer} {et~al.}(2002){Netzer}, {Chelouche}, {George}, {Turner},
  {Crenshaw}, {Kraemer}, \& {Nandra}}]{Net02}
{Netzer}, H., {Chelouche}, D., {George}, I.~M., {et~al.} 2002, \apj, 571, 256,
  \dodoi{10.1086/338967}

\bibitem[{{Noda} \& {Done}(2018)}]{Noda18}
{Noda}, H., \& {Done}, C. 2018, \mnras, 480, 3898,
  \dodoi{10.1093/mnras/sty2032}

\bibitem[{{Noda} {et~al.}(2016){Noda}, {Minezaki}, {Watanabe}, {Kokubo},
  {Kawaguchi}, {Itoh}, {Morihana}, {Saito}, {Nakao}, {Imai}, {Moritani},
  {Takaki}, {Kawabata}, {Nakaoka}, {Uemura}, {Kawabata}, {Yoshida}, {Arai},
  {Takagi}, {Morokuma}, {Doi}, {Itoh}, {Yamada}, {Nakazawa}, {Fukazawa}, \&
  {Makishima}}]{Noda16}
{Noda}, H., {Minezaki}, T., {Watanabe}, M., {et~al.} 2016, \apj, 828, 78,
  \dodoi{10.3847/0004-637X/828/2/78}

\bibitem[{{O'Donnell}(1994)}]{ODo94}
{O'Donnell}, J.~E. 1994, \apj, 422, 158, \dodoi{10.1086/173713}

\bibitem[{{Oknyansky} {et~al.}(2019){Oknyansky}, {Winkler}, {Tsygankov},
  {Lipunov}, {Gorbovskoy}, {van Wyk}, {Buckley}, \& {Tyurina}}]{Okny19}
{Oknyansky}, V.~L., {Winkler}, H., {Tsygankov}, S.~S., {et~al.} 2019, \mnras,
  483, 558, \dodoi{10.1093/mnras/sty3133}

\bibitem[{{Oknyansky} {et~al.}(2021){Oknyansky}, {Brotherton}, {Tsygankov},
  {Dodin}, {Bao}, {Zhao}, {Du}, {Burlak}, {Ikonnikova}, {Tatarnikov},
  {Belinski}, {Fedoteva}, {Shatsky}, {Mishin}, {Zheltouhov}, {Potanin}, {Wang},
  {McLane}, {Kobulnicky}, {Dale}, {Zastrocky}, {Maithil}, {Olson}, {Adelman},
  {Carter}, {Murphree}, {Oeur}, {Schonsberg}, \& {Roth}}]{Okny21}
{Oknyansky}, V.~L., {Brotherton}, M.~S., {Tsygankov}, S.~S., {et~al.} 2021,
  \mnras, 505, 1029, \dodoi{10.1093/mnras/stab1138}

\bibitem[{{Petrucci} {et~al.}(2018){Petrucci}, {Ursini}, {De Rosa}, {Bianchi},
  {Cappi}, {Matt}, {Dadina}, \& {Malzac}}]{Petr18}
{Petrucci}, P.~O., {Ursini}, F., {De Rosa}, A., {et~al.} 2018, \aap, 611, A59,
  \dodoi{10.1051/0004-6361/201731580}

\bibitem[{{Petrucci} {et~al.}(2020){Petrucci}, {Gronkiewicz}, {Rozanska},
  {Belmont}, {Bianchi}, {Czerny}, {Matt}, {Malzac}, {Middei}, {De Rosa},
  {Ursini}, \& {Cappi}}]{Petr20}
{Petrucci}, P.~O., {Gronkiewicz}, D., {Rozanska}, A., {et~al.} 2020, \aap, 634,
  A85, \dodoi{10.1051/0004-6361/201937011}

\bibitem[{{Poole} {et~al.}(2008){Poole}, {Breeveld}, {Page}, {Landsman},
  {Holland}, , {et~al.}}]{Poo08}
{Poole}, T.~S., {Breeveld}, A.~A., {Page}, M.~J., {et~al.} 2008, \mnras, 383,
  627, \dodoi{10.1111/j.1365-2966.2007.12563.x}

\bibitem[{{Porquet} {et~al.}(2018){Porquet}, {Reeves}, {Matt}, {Marinucci},
  {Nardini}, {Braito}, {Lobban}, {Ballantyne}, {Boggs}, {Christensen},
  {Dauser}, {Farrah}, {Garcia}, {Hailey}, {Harrison}, {Stern}, {Tortosa},
  {Ursini}, \& {Zhang}}]{Porq18}
{Porquet}, D., {Reeves}, J.~N., {Matt}, G., {et~al.} 2018, \aap, 609, A42,
  \dodoi{10.1051/0004-6361/201731290}

\bibitem[{{Roming} {et~al.}(2005){Roming}, {Kennedy}, {Mason}, {Nousek}, {Ahr},
  {Bingham}, {Broos}, {Carter}, {Hancock}, {Huckle}, {Hunsberger}, {Kawakami},
  {Killough}, {Koch}, {McLelland}, {Smith}, {Smith}, {Soto}, {Boyd},
  {Breeveld}, {Holland}, {Ivanushkina}, {Pryzby}, {Still}, \& {Stock}}]{Romi05}
{Roming}, P.~W.~A., {Kennedy}, T.~E., {Mason}, K.~O., {et~al.} 2005, \ssr, 120,
  95, \dodoi{10.1007/s11214-005-5095-4}

\bibitem[{{Ruan} {et~al.}(2016){Ruan}, {Anderson}, {Cales}, {Eracleous},
  {Green}, {Morganson}, {Runnoe}, {Shen}, {Wilkinson}, {Blanton}, {Dwelly},
  {Georgakakis}, {Greene}, {LaMassa}, {Merloni}, \& {Schneider}}]{Ruan16}
{Ruan}, J.~J., {Anderson}, S.~F., {Cales}, S.~L., {et~al.} 2016, \apj, 826,
  188, \dodoi{10.3847/0004-637X/826/2/188}

\bibitem[{{Runnoe} {et~al.}(2016){Runnoe}, {Cales}, {Ruan}, {Eracleous},
  {Anderson}, {Shen}, {Green}, {Morganson}, {LaMassa}, {Greene}, {Dwelly},
  {Schneider}, {Merloni}, {Georgakakis}, \& {Roman-Lopes}}]{Runn16}
{Runnoe}, J.~C., {Cales}, S., {Ruan}, J.~J., {et~al.} 2016, \mnras, 455, 1691,
  \dodoi{10.1093/mnras/stv2385}

\bibitem[{{Schlafly} \& {Finkbeiner}(2011)}]{Schl11}
{Schlafly}, E.~F., \& {Finkbeiner}, D.~P. 2011, \apj, 737, 103,
  \dodoi{10.1088/0004-637X/737/2/103}

\bibitem[{{Shapovalova} {et~al.}(2019){Shapovalova}, {Popovi{\'c}}, {},
  {Afanasiev}, {Ili{\'c}}, {}, {Kova{\v{c}}evi{\'c}}, {}, {Burenkov},
  {Chavushyan}, {Mar{\v{c}}eta-Mandi{\'c}}, {}, {Spiridonova}, {Valdes},
  {Bochkarev}, {Pati{\~n}o-{\'A}lvarez}, {Carrasco}, \& {Zhdanova}}]{Shap19}
{Shapovalova}, A.~I., {Popovi{\'c}}, {}, L.~{\v{C}}., {et~al.} 2019, \mnras,
  485, 4790, \dodoi{10.1093/mnras/stz692}

\bibitem[{{Shappee} {et~al.}(2014){Shappee}, {Prieto}, {Grupe}, {Kochanek},
  {Stanek}, {De Rosa}, {Mathur}, {Zu}, {Peterson}, {Pogge}, {Komossa}, {Im},
  {Jencson}, {Holoien}, {Basu}, {Beacom}, {Szczygie{\l}}, {Brimacombe},
  {Adams}, {Campillay}, {Choi}, {Contreras}, {Dietrich}, {Dubberley},
  {Elphick}, {Foale}, {Giustini}, {Gonzalez}, {Hawkins}, {Howell}, {Hsiao},
  {Koss}, {Leighly}, {Morrell}, {Mudd}, {Mullins}, {Nugent}, {Parrent},
  {Phillips}, {Pojmanski}, {Rosing}, {Ross}, {Sand}, {Terndrup}, {Valenti},
  {Walker}, \& {Yoon}}]{Shapp14}
{Shappee}, B.~J., {Prieto}, J.~L., {Grupe}, D., {et~al.} 2014, \apj, 788, 48,
  \dodoi{10.1088/0004-637X/788/1/48}

\bibitem[{{Steenbrugge} {et~al.}(2005){Steenbrugge}, {Kaastra}, {Crenshaw},
  {Kraemer}, {Arav}, {George}, {Liedahl}, {van der Meer}, {Paerels}, {Turner},
  \& {Yaqoob}}]{Stee05}
{Steenbrugge}, K.~C., {Kaastra}, J.~S., {Crenshaw}, D.~M., {et~al.} 2005, \aap,
  434, 569, \dodoi{10.1051/0004-6361:20047138}

\bibitem[{{Trakhtenbrot} {et~al.}(2019){Trakhtenbrot}, {Arcavi}, {MacLeod},
  {Ricci}, {Kara}, {Graham}, {Stern}, {Harrison}, {Burke}, {Hiramatsu},
  {Hosseinzadeh}, {Howell}, {Smartt}, {Rest}, {Prieto}, {Shappee}, {Holoien},
  {Bersier}, {Filippenko}, {Brink}, {Zheng}, {Li}, {Remillard}, \&
  {Loewenstein}}]{Trakh19}
{Trakhtenbrot}, B., {Arcavi}, I., {MacLeod}, C.~L., {et~al.} 2019, \apj, 883,
  94, \dodoi{10.3847/1538-4357/ab39e4}

\bibitem[{{Tran} {et~al.}(1992){Tran}, {Osterbrock}, \& {Martel}}]{Tran92}
{Tran}, H.~D., {Osterbrock}, D.~E., \& {Martel}, A. 1992, \aj, 104, 2072,
  \dodoi{10.1086/116382}

\bibitem[{{Turner} {et~al.}(2011){Turner}, {Miller}, {Kraemer}, \&
  {Reeves}}]{Turn11}
{Turner}, T.~J., {Miller}, L., {Kraemer}, S.~B., \& {Reeves}, J.~N. 2011, \apj,
  733, 48, \dodoi{10.1088/0004-637X/733/1/48}

\bibitem[{{Turner} {et~al.}(2008){Turner}, {Reeves}, {Kraemer}, \&
  {Miller}}]{Tur08}
{Turner}, T.~J., {Reeves}, J.~N., {Kraemer}, S.~B., \& {Miller}, L. 2008, \aap,
  483, 161, \dodoi{10.1051/0004-6361:20078808}

\bibitem[{{Voit} {et~al.}(1987){Voit}, {Shull}, \& {Begelman}}]{Voit87}
{Voit}, G.~M., {Shull}, J.~M., \& {Begelman}, M.~C. 1987, \apj, 316, 573,
  \dodoi{10.1086/165226}

\bibitem[{{Willingale} {et~al.}(2013){Willingale}, {Starling}, {Beardmore},
  {Tanvir}, \& {O'Brien}}]{Will13}
{Willingale}, R., {Starling}, R.~L.~C., {Beardmore}, A.~P., {Tanvir}, N.~R., \&
  {O'Brien}, P.~T. 2013, \mnras, 431, 394, \dodoi{10.1093/mnras/stt175}

\bibitem[{{Yang} {et~al.}(2018){Yang}, {Wu}, {Fan}, {Jiang}, {McGreer},
  {Shangguan}, {Yao}, {Wang}, {Joshi}, {Green}, {Wang}, {Feng}, {Fu}, {Yang},
  \& {Liu}}]{Yang18}
{Yang}, Q., {Wu}, X.-B., {Fan}, X., {et~al.} 2018, \apj, 862, 109,
  \dodoi{10.3847/1538-4357/aaca3a}

\bibitem[{{Zycki} \& {Czerny}(1994)}]{Zyck94}
{Zycki}, P.~T., \& {Czerny}, B. 1994, \mnras, 266, 653

\bibitem[{{Zycki} {et~al.}(1999){Zycki}, {Done}, \& {Smith}}]{Zyck99}
{Zycki}, P.~T., {Done}, C., \& {Smith}, D.~A. 1999, \mnras, 305, 231,
  \dodoi{10.1046/j.1365-8711.1999.02431.x}

\end{thebibliography}
%%%%%%%%%%%%%%%%%%%%%%%%%%%%%%%%%%%%%%%%%%%%%%%%%%%%%%%%%%%%%%%%%%%%%%%%%%%%%%%%%%%%%%%%%%%%%%%%%%%%%%%
%%%%%%%%%%%%%%%%%%%%%%%%%%%%%%%%%%%%%%%%%%%%%%%%%%%%%%%%%%%%%%%%%%%%%%%%%%%%%%%%%%%%%%%%%%%%%%%%%%%%%%%
%%%%%%%%%%%%%%%%%%%%%%%%%%%%%%%%%%%%%%%%%%%%%%%%%%%%%%%%%%%%%%%%%%%%%%%%%%%%%%%%%%%%%%%%%%%%%%%%%%%%%%%
%\appendix

%\section{Appendix section}
%%%%%%%%%%%%%%%%%%%%%%%%%%%%%%%%%%%%%%%%%%%%%%%%%%%%%%%%%%%%%%%%%%%%%%%%%%%%%%%%%%%%%%%%%%%%%%%%%%%%%%%
%%%%%%%%%%%%%%%%%%%%%%%%%%%%%%%%%%%%%%%%%%%%%%%%%%%%%%%%%%%%%%%%%%%%%%%%%%%%%%%%%%%%%%%%%%%%%%%%%%%%%%%
%%%%%%%%%%%%%%%%%%%%%%%%%%%%%%%%%%%%%%%%%%%%%%%%%%%%%%%%%%%%%%%%%%%%%%%%%%%%%%%%%%%%%%%%%%%%%%%%%%%%%%%
\end{document}